\documentclass[numberedappendix,useAMS,usenatbib,letterpaper]{mn2e}
\usepackage[pass]{geometry}
\usepackage{amssymb,amsmath}
\usepackage{verbatim}
\usepackage{color}
\usepackage{hyperref}
\usepackage{booktabs}
\usepackage{needspace}
\usepackage{enumitem}
\usepackage{tikz}
\usepackage{graphicx}
\usetikzlibrary{positioning}
\definecolor{linkcolor}{rgb}{0,0,0.25}
\hypersetup{
  colorlinks=true,        
  linkcolor=linkcolor,    
  citecolor=linkcolor,    
  filecolor=linkcolor,    
  urlcolor=linkcolor      
}
\newcounter{address}
\title[Signatures of resonance and phase mixing]
{Signatures of resonance and phase mixing in the Galactic disc}

\author[J. A. S. Hunt et al.]
{\parbox{\textwidth}{Jason A.~S.~Hunt$^1$, Mathew W. Bub$^2$, Jo Bovy$^{1,2}$, J. Ted Mackereth$^3$, Wilma H. Trick$^4$, and Daisuke Kawata$^5$}\vspace{0.5cm}
\\
$^{1}$ Dunlap Institute for Astronomy and Astrophysics, University of Toronto, 50 St. George Street, Toronto, Ontario, M5S 3H4, Canada\\
$^{2}$ Department of Astronomy and Astrophysics, University of Toronto, 50 St. George Street, Toronto, ON, M5S 3H4, Canada \\
$^{3}$ School of Astronomy and Astrophysics, University of Birmingham, Edgbaston, Birmimgham, B15 2TT, UK\\
$^{4}$ Max-Planck-Insitut f\"{u}r Astrophysik, Karl-Schwarzschild-Str. 1, D-85748 Garching b. M\"{u}nchen, Germany\\
$^5$ Mullard Space Science Laboratory, Holmbury St. Mary, Dorking, RH5 6NT, UK.
}

\pagerange{\pageref{firstpage}--\pageref{lastpage}}
\pubyear{2019}

\begin{document}

\maketitle

\label{firstpage}

\begin{abstract}
$Gaia$ DR2 has provided an unprecedented wealth of information about the kinematics of stars in the Solar neighbourhood, and has highlighted the degree of features in the Galactic disc. We confront the data with a range of bar and spiral models in both action-angle space, and the $R_{\mathrm{G}}-v_{\phi}$ plane. We find that the phase mixing induced by transient spiral structure creates ridges and arches in the local kinematics which are consistent with the $Gaia$ data. We are able to produce a qualitatively good match to the data when combined with a bar with a variety of pattern speeds, and show that it is non trivial to decouple the effects of the bar and the spiral structure.
\end{abstract}


\section{Introduction}\label{intro}
The European Space Agency (ESA)'s $Gaia$ mission \citep{GaiaMission} is transforming our view of the Milky Way. The second data release \citep[DR2;][]{DR2} contains 5 parameter astrometry for around $1.3\times10^9$ stars, allowing us to trace the kinematics of the disc to unprecedented distances. In addition, DR2 also provides radial velocities for around $7.2\times10^6$ stars, enabling a detailed analysis of the Solar neighbourhood \citep[e.g.][]{GCKatz+18}. This exciting new data revealed numerous disequilibria features in the Galactic disc, such as the `ridges' in the $R_{\mathrm{G}}-v_{\phi}$ plane \citep[e.g.][]{KBCCGHS18,Antoja+18} and we now know that the classical moving groups in the $v_{\mathrm{R}}-v_{\phi}$ plane (e.g. Hyades, Pleiades, Coma Berenices, Sirius \& Hercules) are a local manifestation of this radially extended structure. With DR2, \cite{Antoja+18} also discovered the `phase spiral' in the $z-v_z$ plane, highlighting the ongoing phase mixing in the vertical direction. This vertical asymmetry which was first discovered by \cite{WGYDC12} has now been well measured \citep{BB19}, but the discussion of its origin is ongoing.  

One potential explanation for the radial structure is resonant interaction with the Galactic bar, or a fixed pattern speed spiral, although a large number of resonances are required to create every feature. Alternately, \citep{Quillen+18} proposed that the ridges in the Solar neighbourhood could result from individual spiral arm crossings, linking the divisions in the moving groups to specific spiral arms. Another potential explanation for the radial structure is phase wrapping \citep[e.g.][]{MQWFNSB09} either resulting from interaction with a satellite or dwarf galaxy \citep[e.g.][]{LJGG-CB18} or internal dynamical processes such as transient spiral structure \citep[e.g.][]{HHBKG18}. The vertical disequilibria are naturally explained by the external perturbation from a satellite or dwarf galaxy such as Sagittarius \citep[e.g.][]{Antoja+18,LMJG19,Bland-Hawthorn+19}. It has also been proposed that the vertical disequilibra features can arise from secular effects such as the buckling of the Galactic bar \citep{Khoperskov+19} although the similarity of the spiral for stars with different ages \citep{LMJG19} and the lack of a spiral for stars on hotter orbits \citep{LS19} argues against this scenario for its origin. There is likely a combination of both resonance and phase wrapping occurring in the Galactic disc, and it is not trivial to disentangle which feature arises by which specific mechanism, or which combination of mechanisms.

In \cite{HHBKG18} we showed that transient spiral arms combined with a long slow bar reproduce the ridges in the $R_{\mathrm{G}}-v_{\phi}$ plane, as well as producing a qualitatively good match to the Solar neighbourhood kinematics in the $v_{\mathrm{R}}-v_{\phi}$ plane. Recently, \cite{Fragkoudi+19} reproduced the ridge structure with a short fast bar in an $N$-body simulation, and highlighted that these ridges also stand out well when colouring the $R_{\mathrm{G}}-v_{\phi}$ plane by Galactocentric radial velocity. \cite{Khanna+19} showed that both transient spiral structure and an external perturber are able to create these ridges via phase mixing. \cite{M-MPPV19} present a model which reproduces the ridges through the resonances of a bar and density wave spiral arms, and link the ridges with observed `wiggles' in the rotation curve of the model Galaxy, which are similar in amplitude to the wiggles in the rotation curve of the Milky Way, and external galaxies.

\begin{figure*}
\centering
\includegraphics[width=\hsize]{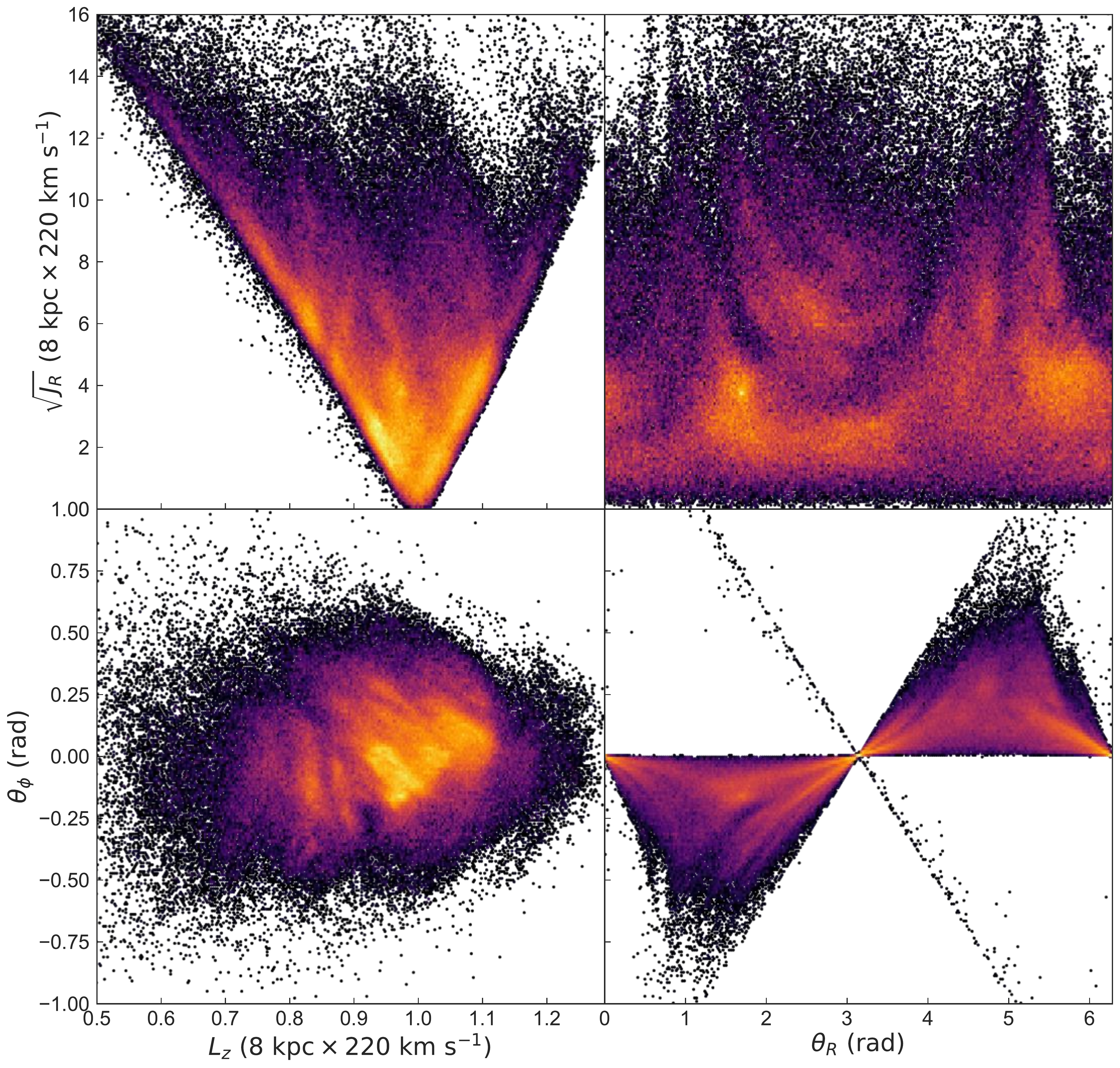}
\caption{Action-angle distribution for $Gaia$ stars with radial velocities, within 200 pc, with fractional parallax errors of less than 10 per cent. The left column shows angular momenta against radial action (upper row) and against azimuthal angle (lower row). The right column shows radial angle against radial action (upper) and azimuthal angle (lower).}
\label{AAGaia}
\end{figure*}

Another way of exploring kinematic structure is through the use of action--angle coordinates. For example, \cite{TCR18} use actions to show that the local moving groups which have long been observed in the Solar neighbourhood $v_{\mathrm{R}}-v_{\phi}$ plane are local, low-$J_{\mathrm{R}}$ manifestations of extended structure in orbit space. They show that while the Hercules, Hyades and Sirius moving groups show
diffusion to higher $J_{\mathrm{R}}$, creating high-$J_{\mathrm{R}}$ ridges at constant slope $\Delta J_{\mathrm{R}}/L_{\mathrm{z}}$, the Pleiades and Coma Berenices moving groups do not, which suggests a potentially different origin. They also point out that the ridges are related to asymmetries in the $v_R$ distribution and low average vertical
action $J_{\mathrm{z}}$.

\cite{Sw+18} explore the effect of various models of spiral structure on the action-angle distribution, specifically a classical density wave model \citep[e.g.][]{LS64}, a model with overlapping transient modes \citep[e.g.][]{SC14}, a swing amplification like model \citep[e.g.][]{T81} and a dressed mass clump model \citep[e.g.][]{TK91}. They compare the features apparent in the action-angle distributions with the data and determine that the data best resembles the transient spiral modes model, and strongly disfavors the dressed mass clump model. They state that the swing amplification model does resemble the data, but the features are broad, whereas the data also contains fine structure, closer resembling the transient mode model. They also argue against the quasi stationary density wave model, but use arguments from previous works as opposed to directly confronting the data.

\begin{figure}
    \centering
    \includegraphics[width=\hsize]{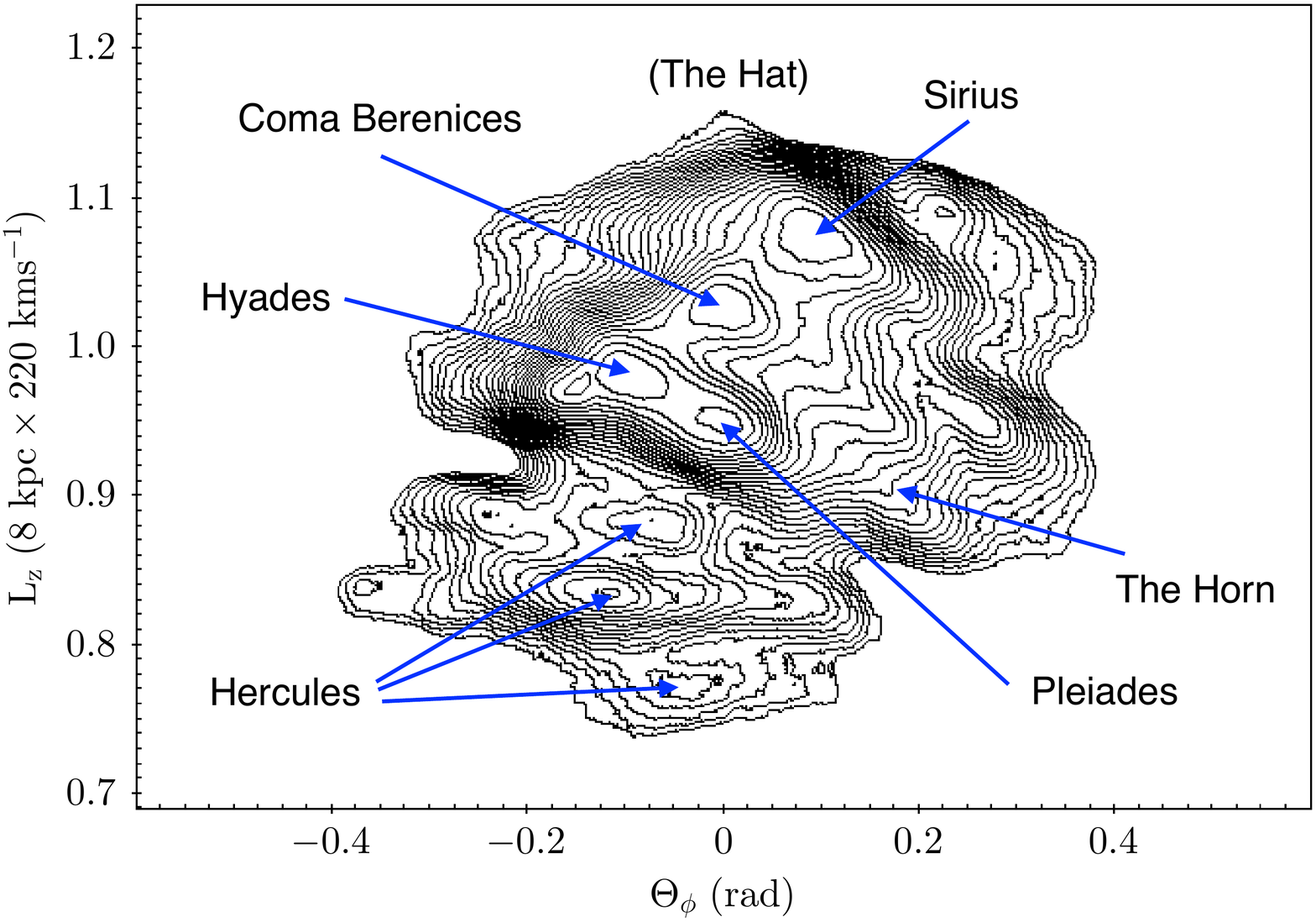}
    \caption{Logarithmic number density contours of the $\theta_{\phi}-L_z$ plane as seen by $Gaia$ DR2, which as expected closely resembles the $v_{\mathrm{R}}-v_{\phi}$ plane. The classical moving groups are labeled along with the horn and the location of the hat, which is below the minimum density for the contours.}
    \label{contours}
\end{figure}
\begin{figure}
    \centering
    \includegraphics[clip, trim=0cm 5cm 0cm 5cm,width=\hsize]{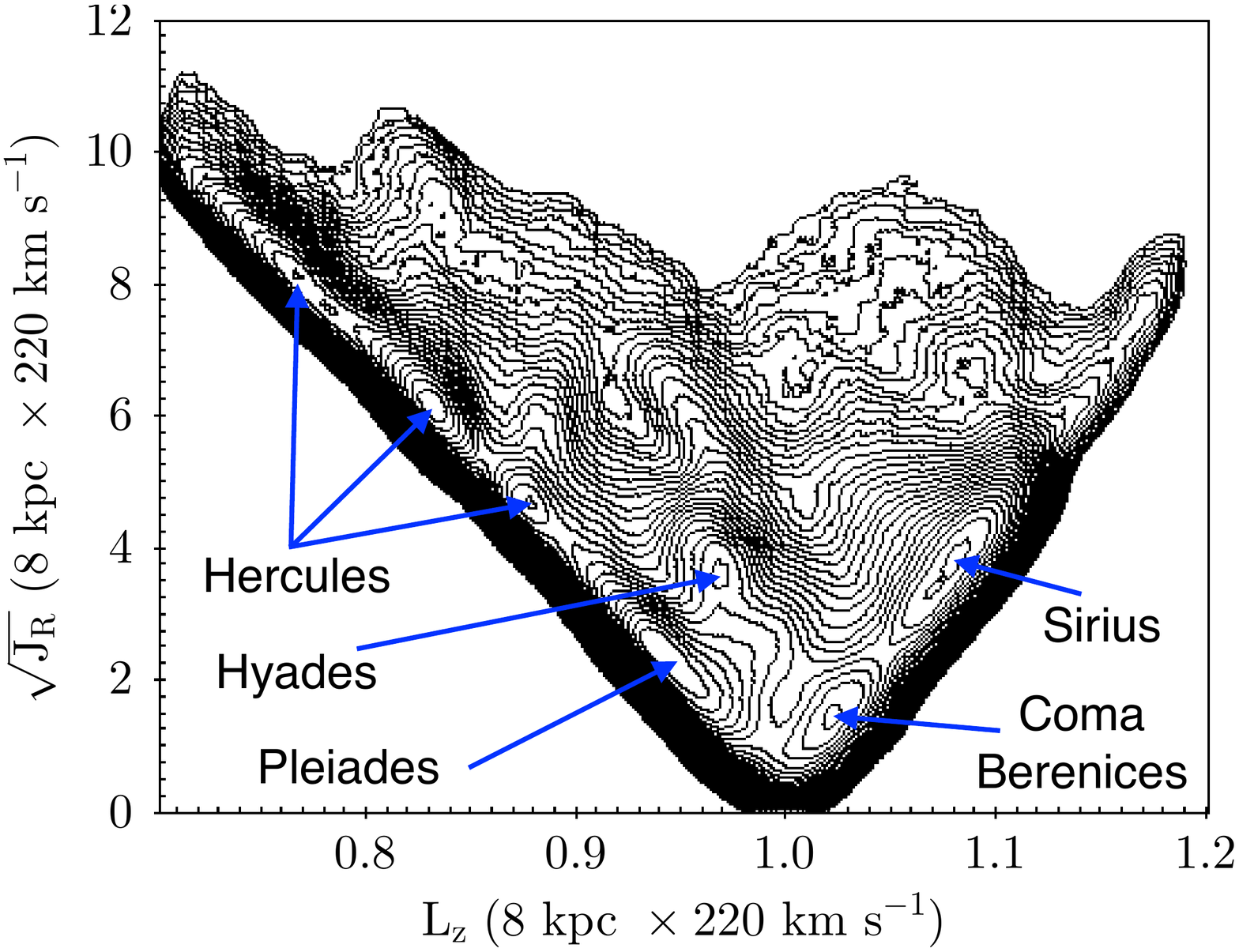}
    \includegraphics[clip, trim=0cm 5cm 0cm 5cm,width=\hsize]{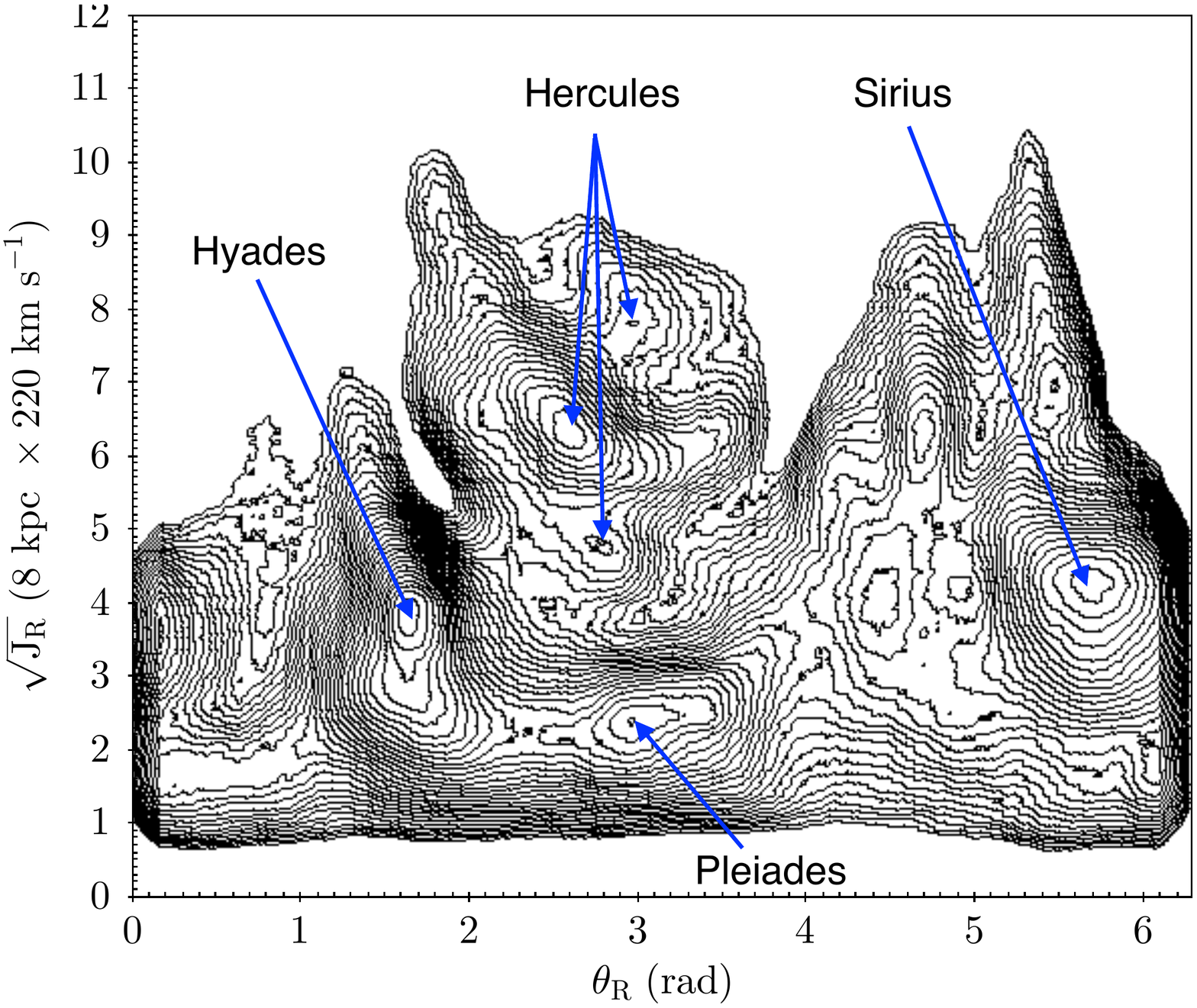}
    \includegraphics[clip, trim=0cm 5cm 0cm 5cm,width=\hsize]{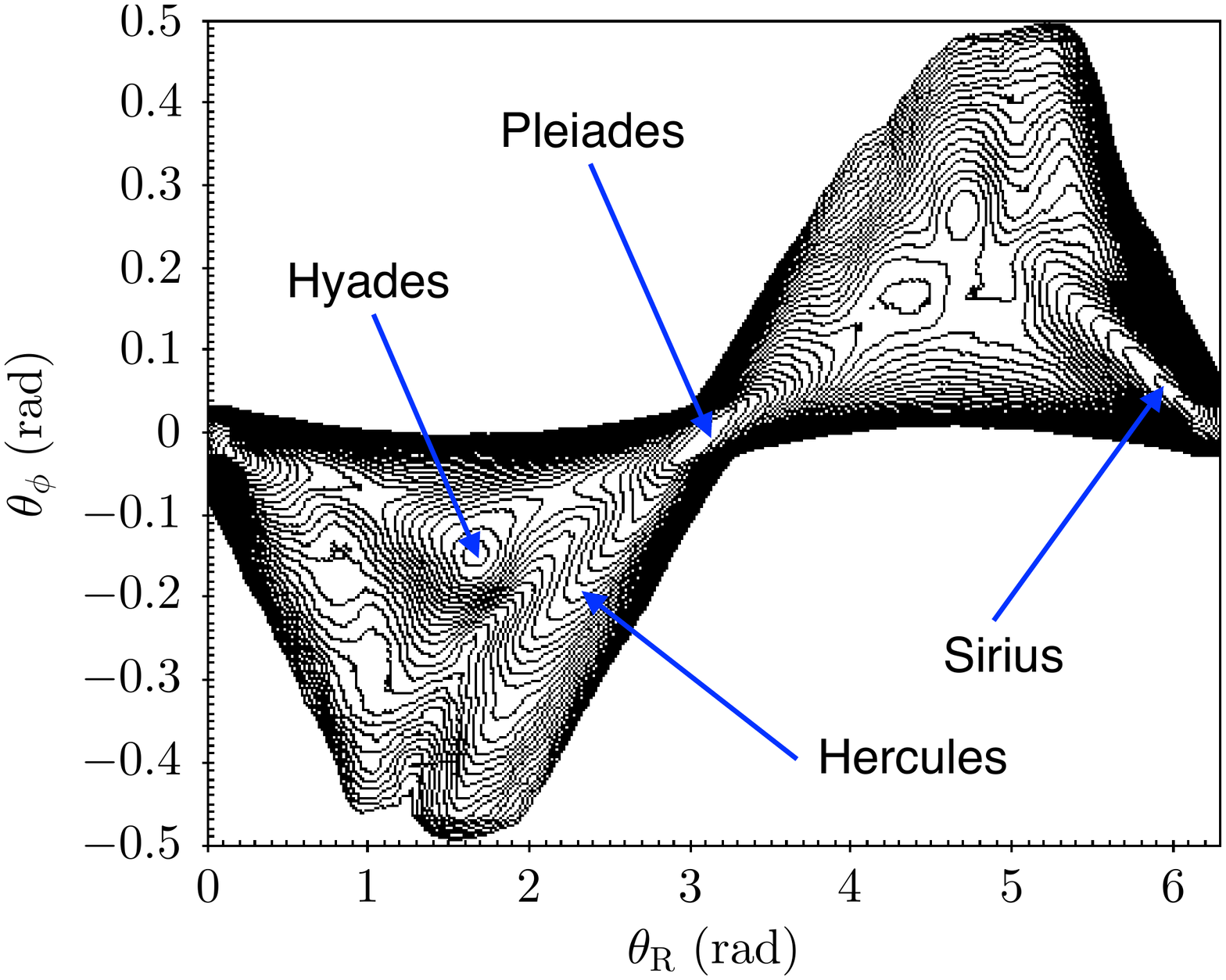}
    \caption{\textbf{Upper:} The $L_z-\sqrt{J_{\mathrm{R}}}$ plane. \textbf{Middle:} The $\theta_{\mathrm{R}}-\sqrt{J_{\mathrm{R}}}$ plane. \textbf{Lower:} The $\theta_{\mathrm{R}}-\theta_{\phi}$ plane, all in logarithmic number density with the classical moving groups labeled when visible.}
    \label{cms}
\end{figure}

\cite{MFSWG18} also examine the action-angle distribution for the long slow bar model of \cite{P-VPWG17}, and find that various resonances fall in the right location to explain features in the Solar neighbourhood kinematics. In this model the Corotation Resonance (CR) causes the lower velocity component of Hercules, the 2:1 Outer Lindblad Resonance (OLR) causes the hat feature, and the 3:1, 4:1 and 6:1 OLRs cause structure in the centre of the velocity distribution near the low velocity moving groups. The importance of considering such higher order bar components was originally demonstrated in \cite{HB18}, where we showed that the 4:1 OLR of the long slow bar model created structure around the low velocity moving groups, and also that for a slightly faster bar the 4:1 OLR would create a Hercules like feature. The strength of Hercules and the hat are not an exact match to the data, and the model does not predict a multiple component Hercules, but other factors like non-axisymmetries such as the spiral arms or an external perturber may account for this. E.g. in \cite{HHBKG18} we showed that the combination of a long bar potential similar to that explored by \cite{MFSWG18} in combination with transient winding spiral arms could reproduce the Solar neighbourhood kinematics, including a double Hercules feature.

In this paper, we explore the effects of the Galactic bar and the spiral structure on the kinematics of stars in the Solar neighbourhood in both the distribution of actions and angles, and the $R_{\mathrm{G}}-v_{\phi}$ plane. In Section \ref{Gaia} we describe the selection and treatment of the $Gaia$ data. In Section \ref{models} we describe the construction of our models and then compare them with the data. Finally, in Section \ref{summary} we summarise the results. 

\section{The Gaia data}
\label{Gaia}
In this section, we describe the selection and treatment of the $Gaia$ data, and the resulting action and angle distribution. We perform a simple selection of all stars with the full six dimensional phase space information which are within 200 pc, with fractional parallax errors of less than 10\%. We calculate distances naively as $d=1/\pi$ which is reliable only for such low parallax errors. We transform the velocities from the $Gaia$ frame to Galactocentric cylindrical coordinates using \texttt{galpy} \citep{B15}, assuming a distance to the Galactic centre of $R_0=8$ kpc, a circular velocity at the Solar radius of $v_{\mathrm{circ}}=220$ km s$^{-1}$ \citep{Betal12}, and a Solar peculiar motion of $v_{\odot}=(U, V, W)=(11.1,12.24,7.25)$ km s$^{-1}$ \citep{SBD10}. While these are fairly standard choices for the above parameters, the effect of these assumptions on the resulting action-angle distribution will be described later.

\subsection{The Solar neighbourhood in action-angle space}
We then calculate the actions, angles and frequencies of the stars using the \texttt{actionAngleStaeckel} \citep{B12-2} function in \texttt{galpy}, assuming the \texttt{MWPotential2014} potential, which is fit to various observational constraints \citep{B15}. We calculate the delta parameter using the \texttt{estimateDeltaStaeckel} routine \citep[e.g. as described in][]{2012MNRAS.426..128S}.

Figure \ref{AAGaia} shows the resulting distribution of planar actions and angles for stars in our sample, similar to that explored in \cite{TCR18} and \cite{Sw+18}. The left column is the azimuthal action $J_{\phi}$, more commonly called angular momentum (around the $z$ axis), $L_z$, and the top row y-axis is radial action (essentially a measure of radial eccentricity), presented as $\sqrt{J_{\mathrm{R}}}$ following \cite{TCR18}. We set the units of the actions to be (8 kpc $\times$ 220 km s$^{-1}$) such that a circular orbit at the assumed Solar position has $L_z=1$. The choice to square root $J_{\mathrm{R}}$ highlights structure at low radial action, which contains the majority of the substructure, in particular the classical moving groups. However, resonant features will lie along a curve for $\sqrt{J_{\mathrm{R}}}$, as opposed to a straight line when visualising $J_{\mathrm{R}}$. The location and slope of such features was already analysed in \cite{TCR18}, and is not a focus of this work. The lower row is the azimuthal angle $\theta_{\phi}$, which is the azimuth of the guiding centre with respect to the Sun. The right column is the radial angle $\theta_{\mathrm{R}}$ which is the phase of the star on its epicycle (with pericenter at $\theta_{\mathrm{R}}=0$, apocenter at $\theta_{\mathrm{R}}=\pi$). The lower left panel of Figure \ref{AAGaia} closely resembles the $v_{\mathrm{R}}-v_{\phi}$ plane which is explored in numerous other works, with clear divisions between the classical moving groups. This is unsurprising as $L_z=R\times v_{\phi}$, and $R_{\mathrm{G}}$ is very similar for all stars in the volume. Similarly, over a small area, $\theta_{\phi}$ should be an approximately linear map to $v_{R}$ \citep[for an illustration, see][]{McMillan2011}. For example, stars with prograde orbits with $0<\theta_{\mathrm{R}}<\pi$ are on the outwards moving part of their epicycle, and have guiding centre azimuths $\theta_{\phi}<0$, and stars with $\theta_{\mathrm{R}}>\pi$ are on the inwards moving part of their epicycle, with guiding centre azimuths $\theta_{\phi}>0$. For example, a star with a larger $\mid\theta_{\phi}\mid$ has its guiding centre azimuth further from the Sun, and  larger $\mid v_{\mathrm{R}}\mid$. This divides the $\theta_{\mathrm{R}}-\theta_{\phi}$ plane clearly into two sections, with the thin diagonal line of stars which does not follow this prescription being on counter rotating orbits. 

Figure \ref{contours} shows the density contour map of the $\theta_{\phi}-L_z$ plane in the standard orientation, with the moving groups labeled. Note that Hercules clearly decomposes into multiple peaks, as shown in other works \citep[e.g.][]{Antoja+18,TCR18,RAF18}. Also note the vertex deviation in the contours such that moving groups at higher $L_z$ occur at higher $\theta_{\phi}$. The Hyades cluster is also visible, as the tiny peak directly to the left of the Hyades moving group. We also label the horn feature \citep[e.g.][]{D00,Fragkoudi+19} and the hat feature which is visible around $L_z=1.2$ in Figure \ref{AAGaia}, but is below the minimum strength for the contour map. Figure \ref{cms} shows contour maps for the other three projections, with the same moving groups labeled. The top panel of Figure \ref{cms} shows the density contour map for the $L_z-\sqrt{J_{\mathrm{R}}}$ plane. The middle panel shows the density contour map for the $\theta_{\mathrm{R}}-\sqrt{J_{\mathrm{R}}}$ plane. Note that Coma Berenices is missing from this projection, because it is located around $\theta_{\mathrm{R}}=0$ (or $2\pi$) and is split across both edges of the plot. The lower panel of Figure \ref{cms} shows the density contours of the $\theta_{\mathrm{R}}-\theta_{\phi}$ plane. In this projection, Hercules does not form distinct peaks, but follows ridges (more clearly visible as three ridges in Figure \ref{AAGaia}), and as with the centre panel Coma Berenices is split across either edge and is not labeled. The hat and the horn do not appear as distinct peaks in these projections. However, there are other obvious structures in the action-angle distributions such as the peaks to the `left' of Sirius in the $\theta_{\mathrm{R}}-\sqrt{J_{\mathrm{R}}}$ and $\theta_{\mathrm{R}}-\theta_{\phi}$ planes around $\theta_{\mathrm{R}}=4.5$, which correspond to the diagonal extension of the contour lines around ($\theta_{\phi},L_z)=(0.2,0.95)$ in Figure \ref{contours} \citep[see][for a more thorough examination of the additional structure in action space]{TCR18}. 

It is not surprising that structure in the Solar
neighbourhood velocity distribution is clear in the action angle
distribution. Over the small area of the sample, all stars essentially have the same $R_{\mathrm{G}}, \phi, z$ coordinates, and the velocities become effectively orbit labels. For spatially more extended survey volumes, actions are the preferred orbit labels. Over a highly localised volume any orbit structure should show up in velocity space, and any velocity structure should show up in orbit space, irrespective of the assumed Galactic potential. So the substructure in Figure \ref{AAGaia} is largely unaffected by the assumption of \texttt{MWPotential2014}. The exact values of the actions and angles will change with a different assupmtion of $R_0, v_{\mathrm{circ}}$, and $v_{\odot}$, but the features will stay qualitatively the same.

\subsection{The resonance criteria}
\label{RC}
As discussed in the introduction, various models exist to explain the substructure in the kinematics of the Solar neighbourhood. A significant fraction propose that the substructure arises from various resonances of the Galactic bar or spiral structure. Assuming that this is the case, then we should be able to fit the overdensities and gaps in the action-angle distribution to specific resonances, arising from a rigidly rotating structure, such as is done in \cite{MFSWG18}.

\begin{figure}
\centering
\includegraphics[width=\hsize]{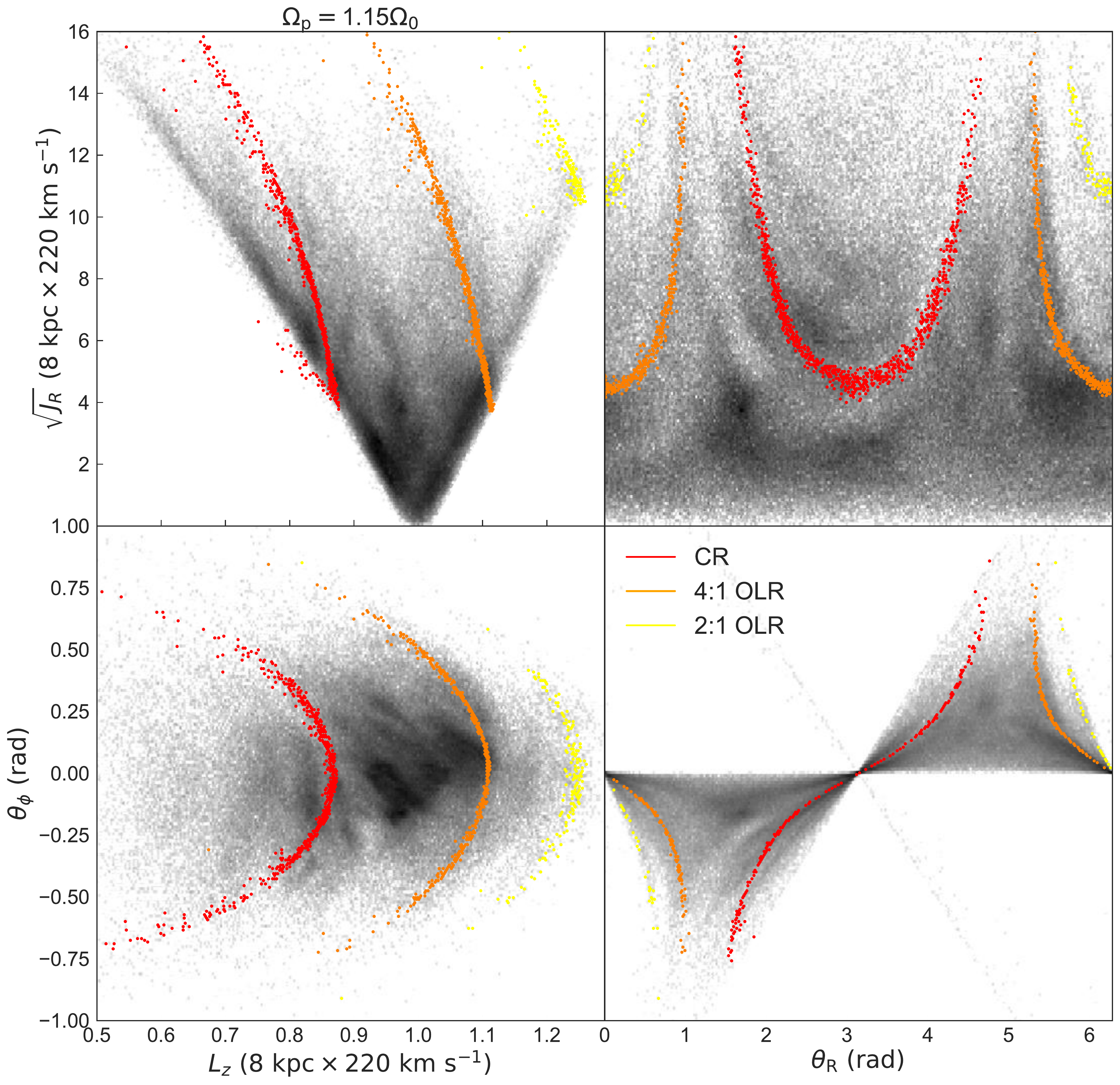}
\caption{Same as Figure \ref{AAGaia}, except overlaid with the CR (red), the 4:1 OLR (orange) and the 2:1 OLR (yellow) for a rigidly rotating pattern of $\Omega_{\mathrm{p}}=1.15\times\Omega_0$.}
\label{RCrit}
\end{figure}

The frequencies which are calculated alongside the actions and angles enable us to compute the `resonance criteria', for which stars with a specific action and angle (and assuming a Milky Way potential) are affected by such a resonance. For example, the corotation resonance for a rigidly rotating structure with pattern speed $\Omega_{\mathrm{p}}$ occurs when $\Omega_{\mathrm{p}}-\Omega_{\phi}=0$. The inner (ILR) and outer (OLR) 2:1 Lindblad resonances occur when $\Omega_{\mathrm{p}}-\Omega_{\phi}\pm\Omega_{\mathrm{R}}/2=0$, and similarly for the 4:1 ILR and OLR, when $\Omega_{\mathrm{p}}-\Omega_{\phi}\pm\Omega_{\mathrm{R}}/4=0$, and the 1:1 ILR and OLR\ when $\Omega_{\mathrm{p}}-\Omega_{\phi}\pm\Omega_{\mathrm{R}}=0$.

Figure \ref{RCrit} shows the action angle distribution from Figure \ref{AAGaia} overlaid with the stars which are close to the resonance criteria for the CR (red), the 4:1 OLR (orange) and the 2:1 OLR (yellow), when assuming a rigidly rotating structure with a pattern speed of $\Omega_{\mathrm{p}}=1.15\times\Omega_0$, where we assume a local circular frequency of $\Omega_0=27.5$ km s$^{-1}$ kpc$^{-1}$. We are not claiming this is a best fitting model for the Milky Way, merely using it as an illustration of the axisymmetric resonance lines, and the similarity between the slope of the resonance lines and the gaps and overdensities in the data.

In this simple model, the 2:1 OLR (yellow) lies in the high angular momenta region of orbit space, and could be responsible for the faint arch feature found there. The 4:1 OLR (orange) corresponds roughly to the sharp transition around $L_z\sim1.1$ and the CR (red) falls roughly in line with Hercules. However, such a correlation does not imply causation, and a different choice of pattern speed will align different resonances with different features. In addition, the resonance criteria can only be calculated for the axisymmetric potential used to calculate the actions, angles and frequencies, and thus while the resonance lines are symmetric in angle, this is not the case in the $Gaia$ data. In addition, calculations of frequency are affected by the choice of potential which we will illustrate in a later section. Thus, because the assumed potential affects the location of the resonance criteria lines, and the choice of Solar peculiar motion affects the action-angle distribution directly, the combination of effects make it very challenging to directly measure the pattern speed of the structure causing gaps or overdensities from such an analysis alone, even if we assume that such features arise from a single rigidly rotating structure such as the Galactic bar, which may not be the case.

\section{The model}\label{models}
To explore the effects of calculating actions for a non-axisymmetric system, using an axisymmetric potential, and the assumption of a potential, we create two dimensional models using the backwards integration technique of \cite{D00} \citep[e.g. following][]{HB18,HHBKG18}. We transform them into action angle space by sampling $10^7$ phase-space points from the resulting distribution and converting these to action-angle space using the \texttt{actionAngelSpherical} function with the \texttt{LogarithmicHaloPotential}.

As was done in \cite{Hunt+18}, \cite{HB18} and \cite{HHBKG18} we use a Dehnen distribution function \citep{Dehnen99}, which is a function of energy, $E$, and angular momentum, $L$, to model the stellar disc before bar and spiral formation. We reproduce the equations here for convenience. The Dehnen distribution function represents the distribution of stellar orbits such that
\begin{equation}
f_{\text{dehnen}}(E,L)\ \propto \frac{\Sigma(R_e)}{\sigma^2_{\text{R}}(R_e)}\exp\biggl[\frac{\Omega(R_e)[L-L_c(E)]}{\sigma^2_{\text{R}}(R_e)}\biggr],
\label{DDF}
\end{equation}
where $R_e$, $\Omega(R_e)$, and $L_c$ are the radius, angular frequency and angular momentum, respectively, of a circular orbit with energy $E$. We assume a simple power law for the gravitational potential such that the circular velocity is given by
\begin{equation}
  v_c(R)=v_0(R/R_0)^{\beta}\,,
\end{equation}
where $v_0$ is the circular velocity at the solar circle at radius $R_0$.

For all subsequent bar models we use the general form of the $\cos(m\phi)$ potential shown in \cite{HB18}, adapted from the quadrupole potential from \citet{D00}. In \texttt{galpy}, this is the \texttt{CosmphiDiskPotential} model. The bar potential is given by
\begin{equation}
\begin{split}
&\Phi_{\mathrm{b}}(R,\phi)=A_{\text{b}}(t)\cos(m(\phi-\phi_{\text{b}}t))\\ 
& \quad \quad \times
\left\{ \begin{array}{ll} -(R/R_0)^p, & \mathrm{for}\ R \geq R_{\text{b}},\\ ([R_{\text{b}}/R]^p-2)\times(R_{\mathrm{b}}/R_0)^p, & \mathrm{for}\ R \leq R_{\text{b}}, \end{array}
\right.
\end{split}
\end{equation}
where $\phi_{\mathrm{b}}$ is the angle of the bar with respect to the Sun--Galactic-center line, and the bar radius, $R_{\text{b}}$, is set to $80\%$ of the corotation radius. The potential is equivalent to the \cite{D00} quadrupole bar for $m=2$ and $p=-3$, where $m$ is the integer multiple of the $\cos$ term, and $p$ is the power law index.

The bar is grown smoothly using \texttt{galpy}'s  \texttt{DehnenSmoothWrapperPotential} such that
\begin{eqnarray}
A_{\text{b}}(t)=
\left\{\begin{array}{ll} 0,\ \frac{t}{T_{\text{b}}}<t_{\text{1}} \\ A_f\biggl[\frac{3}{16}\xi^5-\frac{5}{8}\xi^3+\frac{15}{16}\xi+\frac{1}{2}\biggr], t_{\text{1}}\leq\frac{t}{T_{\text{b}}}\leq t_{\text{1}}+t_{\text{2}}, \\ A_f,\ \frac{t}{T_{\text{b}}} > t_{\text{1}} + t_{\text{2}}.  \end{array}
\right.\,
\end{eqnarray}
where $t_1$ is the start of bar growth, set to half the integration time, $t_2$ is the duration of the bar growth and $T_{\text{b}}=2\pi/\Omega_{\text{b}}$ is the bar period such that
\begin{equation}
\xi=2\frac{t/T_{\text{b}}-t_{\text{1}}}{t_{\text{2}}}-1,
\end{equation}
and
\begin{equation}
A_f=\alpha_{m}\frac{v_0^2}{p}\biggl(\frac{R_0}{R_b}\biggr)^{p},
\end{equation}
where $\alpha_{m}$ is the dimensionless ratio of forces owing to the $\cos(m\phi)$ component of the bar potential and the axisymmetric background potential, $\Phi_0$, at Galactocentric radius $R_0$ along the bar's major axis. This growth mechanism ensures that the bar amplitude along with it's first and second derivatives are continuous for all $t$, allowing a smooth transition from the non-barred to barred state \citep{D00}.

For our spiral arm potential, both density wave and corotating, we use the \texttt{SpiralArmsPotential} from \texttt{galpy}, which is an implementation of the sinusoidal potential from \cite{CG02} such that
\begin{eqnarray}
\label{SAP}
\Phi(R,\phi,z)&=&-4\pi GH\rho_0\exp\biggl(\frac{r_{\mathrm{0}}-R}{R_s}\biggr)  \nonumber \\
&\times&\sum\frac{C_n}{K_n D_n}\cos(n\gamma)\biggl[\mathrm{sech}\biggl(\frac{K_n z}{\beta_n}\biggr)\biggr]^{B_n},
\end{eqnarray}
where
\begin{equation}
K_n=\frac{nN}{R\sin(\theta_{\mathrm{sp}})},
\end{equation}
\begin{equation}
B_n=K_nH(1+0.4K_nH),
\end{equation}
\begin{equation}
D_n=\frac{1+K_nH+0.3(K_nH)^2}{1+0.3K_nH},
\end{equation}
\begin{equation}
\gamma=N\biggl[\phi-\phi_{\mathrm{ref}}-\frac{\ln(R/r_{\mathrm{0}})}{\tan(\theta_{\mathrm{sp}})}\biggr],
\end{equation}
$N$ is the number of spiral arms, $\theta_{\mathrm{sp}}$ is the pitch angle, $\rho_0$ is the density at $r_0$, $\phi_{\mathrm{ref}}$ is the reference angle, $R_s$ is the radial scale length of the arm and $H$ is the scale height of the arm. Setting $C_n$ to 1 gives a purely sinusoidal potential profile. Alternatively, setting $C_n=[8/3\pi,1/2,8/15\pi]$ results in a potential which behaves approximately as a cosine squared in the arms, and is flat in the inter-arm region \citep{CG02}. Note that while Equation (\ref{SAP}) gives the full form available in \texttt{galpy}, we use the planar form $\Phi(R,\phi,z=0)$, which sets the sech term to 1.

For the quasi-stationary density wave like spiral arms, we assign a fixed pattern speed $\Omega_{\mathrm{sp}}$, and grow the potential in the same fashion as the bar. In this work we also grow the spiral over the same timescale.

For the corotatating, winding spiral potential, we wrap the \texttt{SpiralArmsPotential} from Equation (\ref{SAP}) in \texttt{galpy}'s \texttt{CorotatingRotationWrapperPotential}, such that
\begin{equation}
\phi \rightarrow \phi + \frac{V_p(R)}{R} \times \left(t-t_0\right) + a_{\mathrm{p}}
\end{equation}
and
\begin{equation}
V_p(R) = V_{p,0}\,\left(\frac{R}{R_0}\right)^\beta\,,
\end{equation}
where $V_p(R)$ is the circular velocity curve, $t_0$ is the time when the potential is unchanged by the wrapper and $a_{\mathrm{p}}$ is the position angle at time $t_0$. This causes the arm to wind up over time, as seen in $N$-body simulations. This model is designed to mimic the transient recurrent arms which corotate with the stars at all radii, e.g. as described in \cite{GKC12} and \cite{HHBKG18}. It is not designed to reproduce the classic picture of swing amplification \citep[e.g.][]{T81}.

We then weight the amplitude with a Gaussian using the \texttt{GaussianAmplitudeWrapperPotential} to control the strength of the transient arm, where the amplitude gets multiplied with the function
\begin{equation}
A(t) = \exp\left(-\frac{[t-t_0]^2}{2\,\sigma^2}\right),
\end{equation}
and $\sigma$ is the standard deviation of the Gaussian, which controls the lifetime of the transient spiral potential. Although the wings of the Gaussian technically stretch to infinity, the density enhancements last approximately $\mathcal{L}\approx5.6\times\sigma$ from formation to disruption. This is a simple model potential to approximate the co-rotating recurrent arms observed in $N$-body simulations.

\begin{figure}
\centering
\includegraphics[width=\hsize]{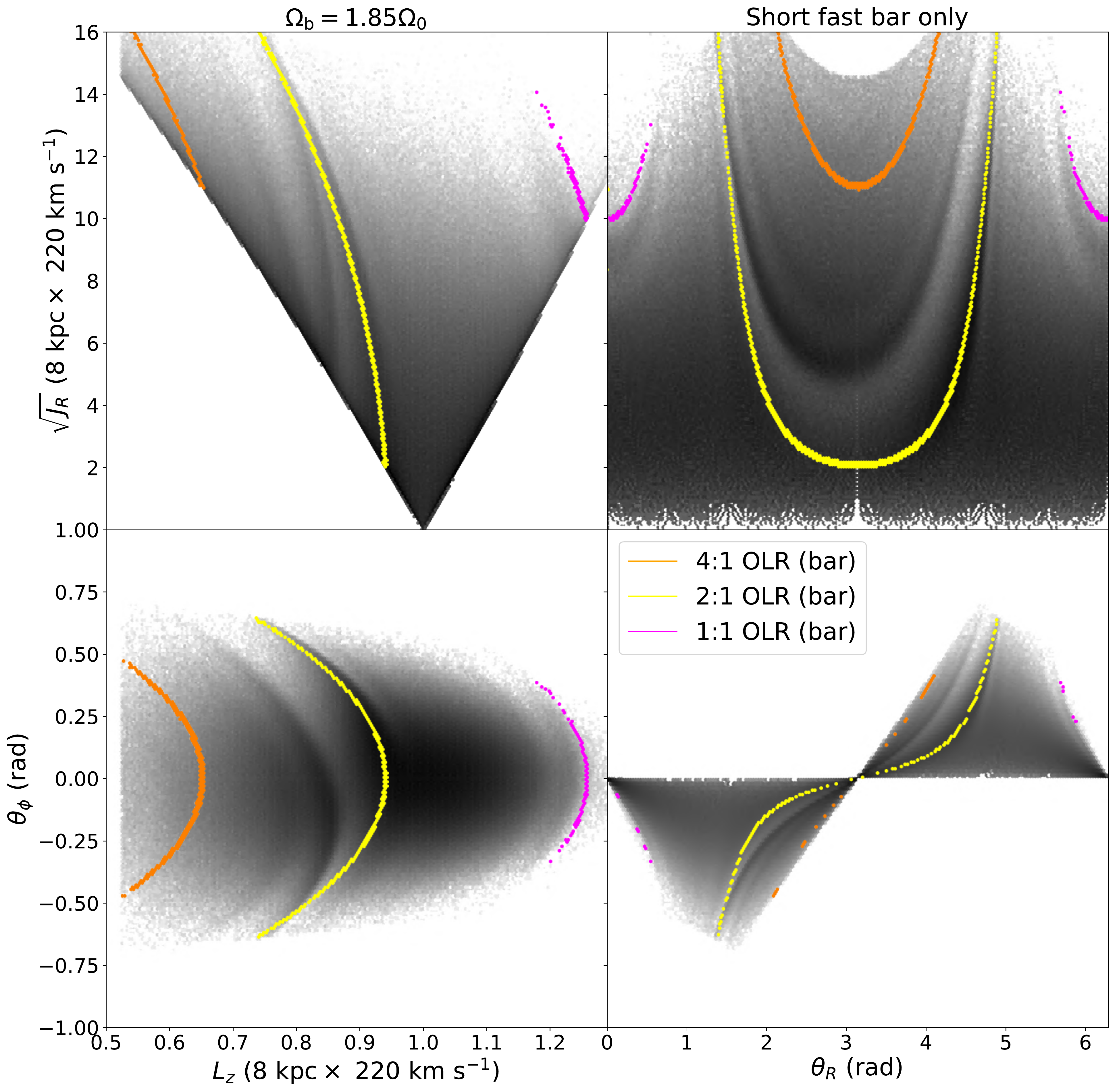}
\caption{Action-angle distribution for Model A, the short-fast bar of \citet{D00}, with $R_b=3.5$ kpc, and $\Omega_{\mathrm{b}}=1.85\times\Omega_0$. The colored lines show the location of the 4:1 OLR (orange), the 2:1 OLR (yellow) and the 1:1 OLR (magenta).}
\label{1p85}
\end{figure}

\begin{table*}
\caption{Model parameters for Models A-H, including the bar half length, $R_{\mathrm{b}}$ (kpc), the bar pattern speed, $\Omega_{\mathrm{b}}\times\Omega_0$, the spiral pattern speed $\Omega_{\mathrm{sp}}\times\Omega_0$ and the origin of the model if applicable.}
\begin{tabular}{@{}lllllll@{}}
\toprule
Label & $R_{\mathrm{b}}$ (kpc) & $\Omega_{\mathrm{b}}$ ($\times\Omega_0$) & $\alpha_2$ & $\alpha_4$ & $\Omega_{\mathrm{sp}}$ ($\times\Omega_0$) & Model \\ \midrule
A & 3.5 & 1.85 & 0.01 & 0 & None & \cite{D00} \\
B & 5.0 & 1.3 & 0.024 & 0.001 & None & \cite{P-VPWG17} \\
C & 5.0 & 1.4 & 0.01 & 0.0005 & None & \cite{HB18} \\
D & 3.5 & 1.85 & 0.01 & 0 & 0.56 \\
E & 3.2 & 1.82 & 0.01 & 0 & 0.92 & \cite{M-MPPV19} \\
F & 5.0 & 1.3 & 0.01 & 0.0005 & Winding  & \cite{HHBKG18} \\
G & 3.8 & 1.6 & 0.01 & 0.0005 & Winding  \\
H & 3.5 & 1.85 & 0.01 & 0.0005 & Winding &  \\ \bottomrule
\end{tabular}
\label{params}
\end{table*}

\begin{figure}
\centering
\includegraphics[width=\hsize]{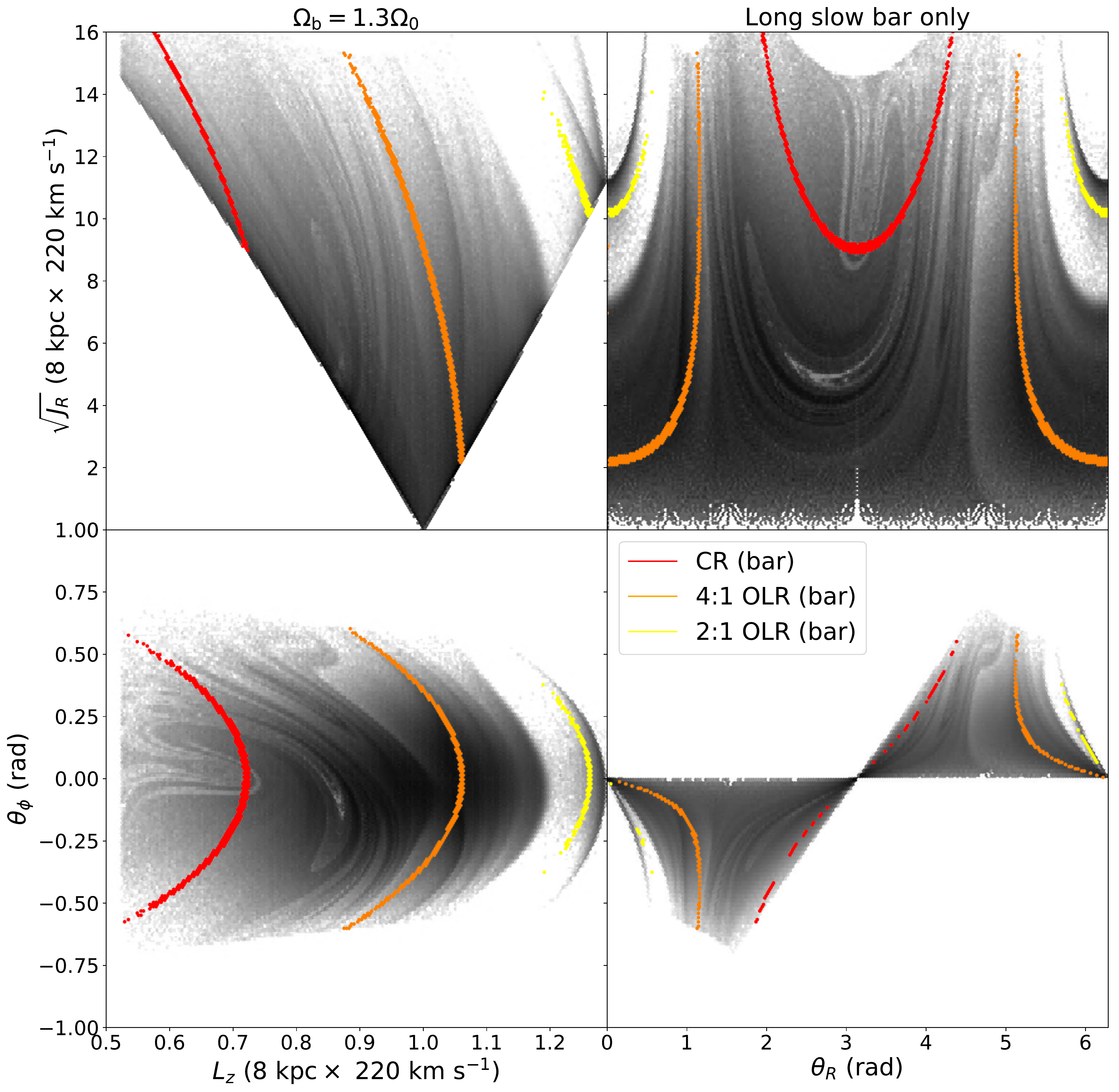}
\caption{Action-angle distribution for Model B, the long-slow bar resembling \citet{P-VPWG17}, with $R_b=5$ kpc, and $\Omega_{\mathrm{b}}=1.3\times\Omega_0$. The coloured lines show the location of the CR (red), the 4:1 OLR (orange) and the 2:1 OLR (yellow).}
\label{1p3}
\end{figure}

\section{Model results}
\subsection{Bar only}
As an initial illustration we present the action angle distribution for a few bar models from the literature which have been claimed to reproduce the Hercules stream, i.e. a short fast bar \citep[from][]{D00}, a long slow bar \citep[similar to][]{P-VPWG17} and a slightly faster long bar \citep[from][]{HB18}. The parameters for these models, and those in subsequent sections are given in Table \ref{params}.

Figure \ref{1p85} shows the action-angle distribution for Model A, a short fast bar, which has been proposed to reproduce the Hercules stream through the OLR \citep[e.g.][]{D00,Hunt+18}. The model parameters used here are $R_{\mathrm{b}}=3.5$ kpc, $\Omega_{\mathrm{b}}=1.85\times\Omega_0$, $\alpha_{\mathrm{2}}=0.01$, $\alpha_{4}=0$, $R_0=8$ kpc and $v_0=220$. The gap and overdensity in the distribution corresponding to the OLR (yellow) is clear in each panel. Very little structure exists in the rest of the distribution, including the area around the 4:1 OLR (orange), which is unsurprising as this simple bar model consists of a purely $m=2$ bar. The exact shape of the Hercules like feature does not perfectly match that of the $Gaia$ data, although a better fit can likely be constructed by altering the bar angle, or other parameters \citep[e.g. see][for an exploration of the model parameter space]{D00}. Interestingly, even this simple model creates three ridges in the $L_z-J_{\mathrm{R}}$ plane, which are qualitatively consistent with the data. The rightmost ridge which lies exactly under the yellow OLR line corresponds to the horn like feature, then slightly to its left is a ridge with very little slope corresponding to the strongest ridge of the Hercules like feature, and then another ridge to its left with a higher slope. The ability for this simple model to reproduce a multiple component Hercules feature is not surprising, smaller secondary peaks are visible in some of the models in \cite{D00}. There is a slight hat feature resulting from the 1:1 OLR (magenta). In general, this model has been well explored in the literature \citep[e.g.][]{Bovy10,Aetal14-2,Monari+16}, and comfortably reproduces the Hercules stream, but none of the other moving groups, or features such as the arches. This is not new, merely shown for comparison.

\begin{figure}
\centering
\includegraphics[width=\hsize]{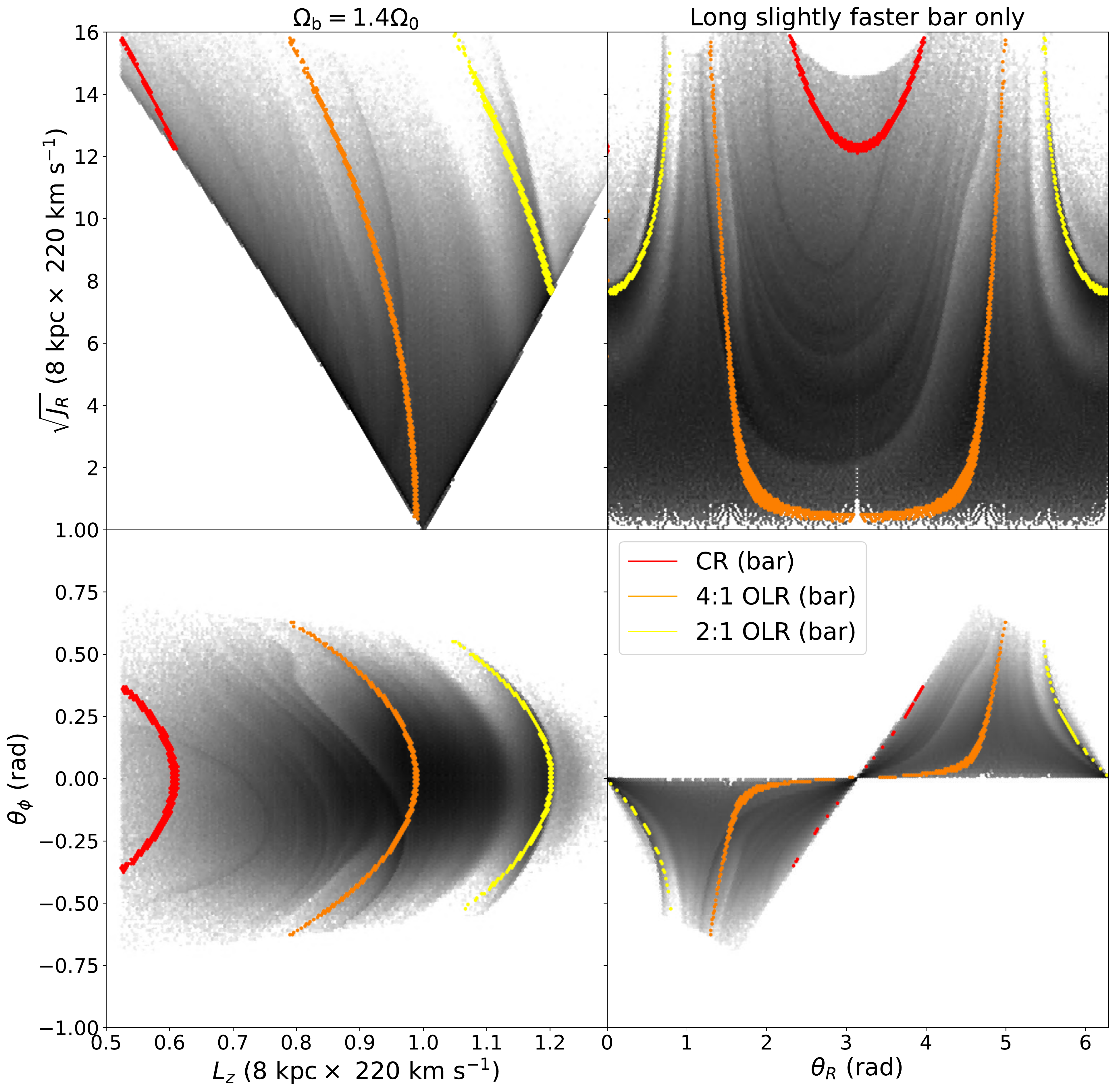}
\includegraphics[width=\hsize]{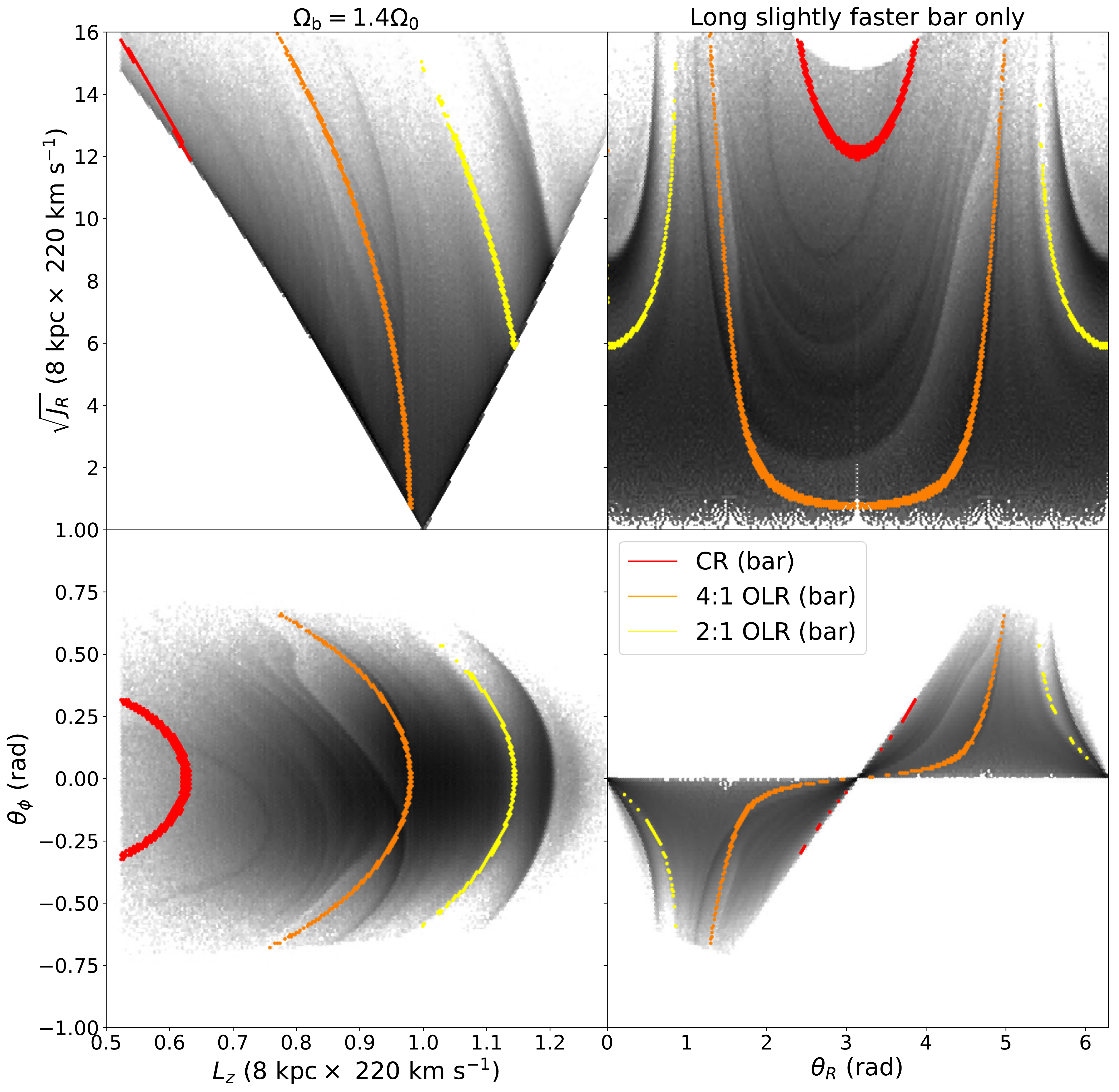}
\caption{Action-angle distribution for Model C, the long-slow $m=4$ bar of \citet{HB18}, with $R_b=5$ kpc, and $\Omega_{\mathrm{b}}=1.4\times\Omega_0$. The colored lines show the location of the 2:1 OLR (yellow), the 4:1 OLR (orange), and the CR (red) for actions, angles and frequencies calculated in the correct axisymmetric potential (upper) and a different axisymmetric potential with the wrong rotation curve (lower).}
\label{1p4}
\end{figure}

\begin{figure}
\centering
\includegraphics[width=\hsize]{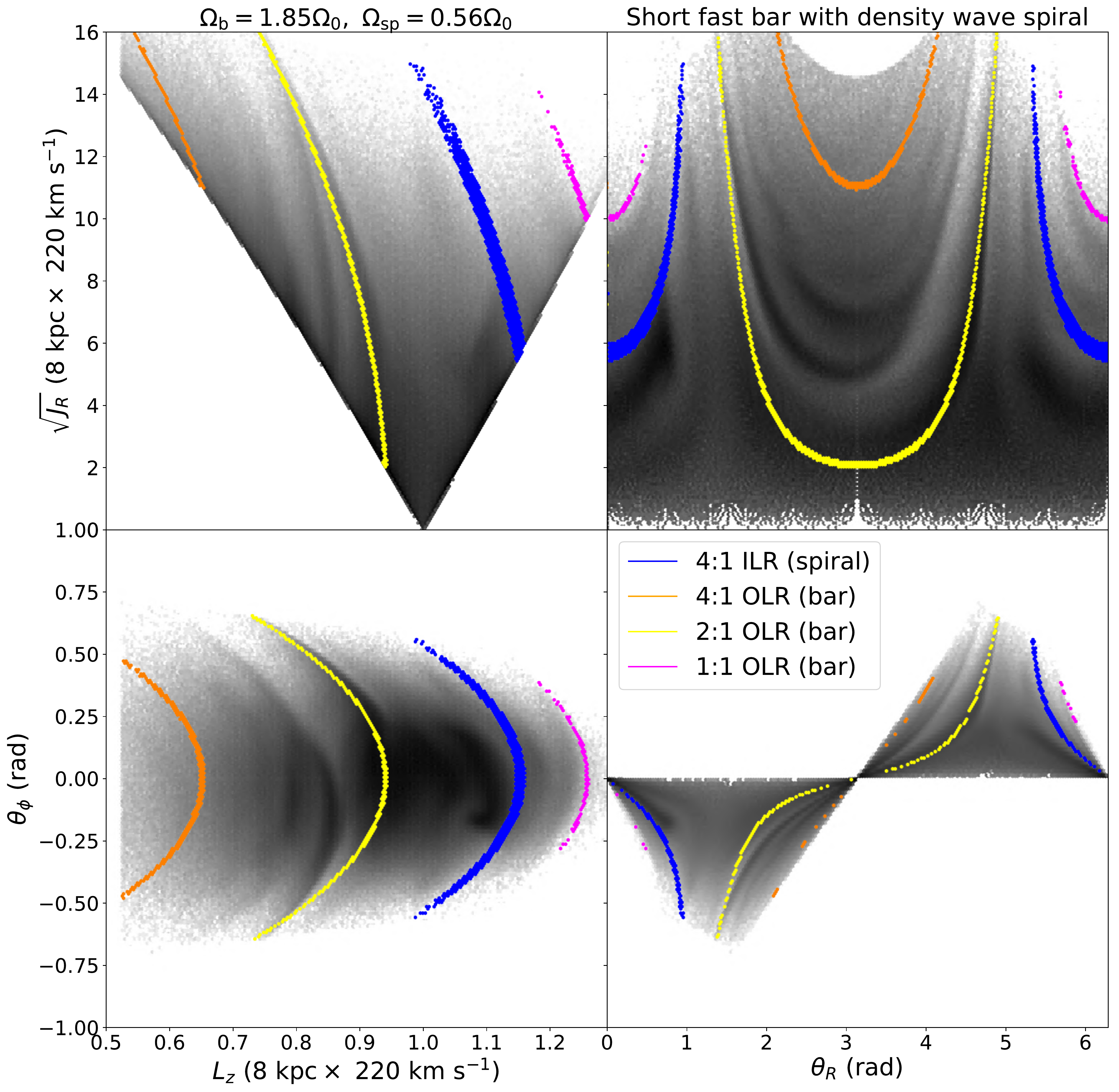}
\includegraphics[width=\hsize]{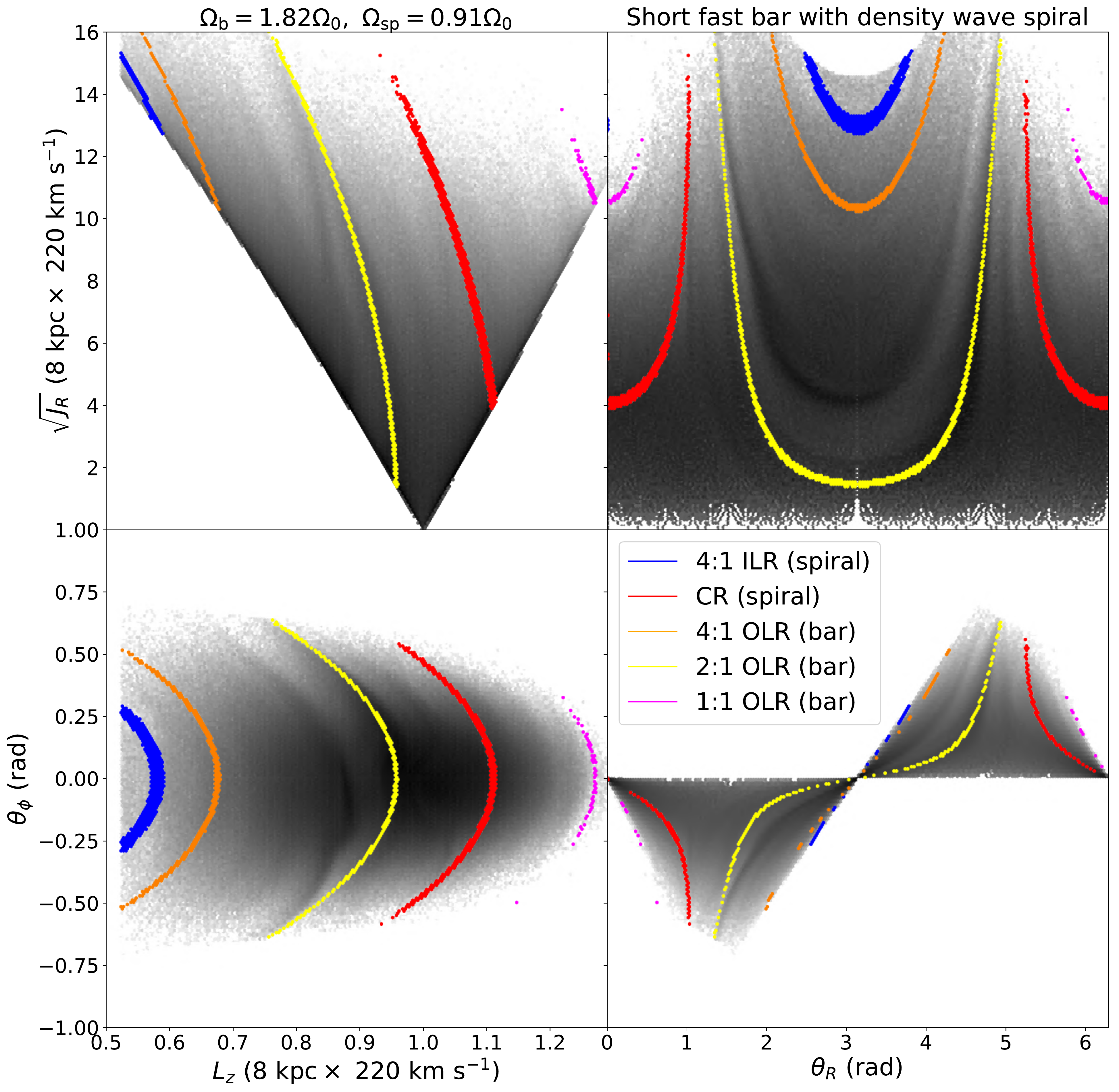}
\caption{Action-angle distribution for Model D, the short-fast bar of \citet{D00}, with $R_b=3.5$ kpc, and $\Omega_{\mathrm{b}}=1.85\times\Omega_0$ combined with a $N=4$ density wave spiral with pattern speed $\Omega_{\mathrm{sp}}=0.56\times\Omega_0$ (upper) and Model E, a short-fast bar, with $R_b=3.2$ kpc, and $\Omega_{\mathrm{b}}=1.82\times\Omega_0$ combined with a $N=4$ density wave spiral with pattern speed $\Omega_{\mathrm{sp}}=0.91\times\Omega_0$ (lower). The colored lines show the location of the bar 2:1 OLR (yellow), the bar 4:1 OLR (orange), the spiral CR (red) and the spiral 4:1 ILR (blue).}
\label{1p8DW4}
\end{figure}

Figure \ref{1p3} shows the action-angle distribution for Model B, a long-slow bar, which has been proposed to reproduce the Hercules stream through the CR \citep[e.g.][]{P-VPWG17}, Note that we are not exploring the model of \cite{P-VPWG17} specifically, \citep[which was recently done in][]{MFSWG18}, but the major features of a bar with this pattern speed should be qualitatively the same. The model parameters used here are $R_{\mathrm{b}}=5$ kpc, $\Omega_{\mathrm{b}}=1.3\times\Omega_0$, $\alpha_{\mathrm{2}}=0.024$, $\alpha_{4}=-0.001$, $R_0=8.4$ kpc and $v_0=242$. The model creates a hat like feature at high $L_z$ with the bar's OLR (yellow). As with other works \citep[e.g.][]{HB18} we find that the CR (red) has only a slight effect on local kinematics, even when including a higher bar mass. As shown in \cite{MFSWG18}, the resonance criteria lines fall in the correct locations to explain features in the velocity distribution, but as with other studies \cite[e.g.][]{Hunt+18,Binney2018} the Hercules like feature is weaker than observed, and the hat like feature is stronger than observed. However, the addition of the $m=4$ bar component to the model \cite[as proposed in][]{HB18} produces an excellent qualitative reproduction of the division between the Sirius and Coma Berenices like streams in the $L_z-\theta_{\phi}$ plane, including its shape and angle, and qualitatively similar in the $\theta_{\mathrm{R}}-J_{\mathrm{R}}$ plane. Also, while the CR does not make a strong Hercules feature alone, it's been shown that the addition of spiral structure in combination with the CR can create a strong distinct Hercules \citep{HHBKG18}, and \cite{MFSWG18} show that the $m=3$ and $m=6$ resonances also create structure for the model of \cite{P-VPWG17}.

The top panel of Figure \ref{1p4} shows the action-angle distribution for Model C, a slightly faster long-slow bar with an $m=4$ Fourier component which has been proposed to reproduce the Hercules stream through the 4:1 OLR \citep{HB18}. The model parameters used here are $R_{\mathrm{b}}=5$ kpc, $\Omega_{\mathrm{b}}=1.4\times\Omega_0$, $\alpha_{\mathrm{2}}=0.01$, $\alpha_{4}=-0.0005$, $R_0=8$ kpc and $v_0=220$. The 4:1 OLR (orange) creates a Hercules like feature in the distribution, which also has a similar angle to the gap in the $Gaia$ data, but is weaker than observed. In addition, the 2:1 OLR (yellow) creates a distinct hat feature, which is qualitatively similar to the $Gaia$ data, but is too strong and occurs at too low $L_z$. The lower panel of Figure \ref{1p4} shows the same model, C, where the actions, angles and frequencies are calculated using the wrong axisymmetric potential, namely \texttt{MWPotential2014} from \texttt{galpy}, instead of the \texttt{LogarithmicHaloPotential} which was used to construct the model. \texttt{MWPotential2014} has a falling rotation curve, and while the action angle distribution barely changes, the resonance criteria lines which are based on the frequencies do shift. The effect on the CR (red) and 4:1 OLR (orange) is relatively small, but the OLR (yellow) shows a significant shift between panels. This is as expected because the falling rotation curve will be more different from the flat rotation curve of the \texttt{LogarithmicHaloPotential} at larger Galactic radii. 

All three bar models can qualitatively reproduce one or more of the features in the local velocity distribution, but none of them can reproduce all the features of the observed Solar neighbourhood kinematics. For example, the short fast bar model \citep[e.g.][]{D00} produces a good reproduction of Hercules but nothing else. The long slow bar model of \cite{P-VPWG17} produces good fits to the hat and low velocity moving groups but does not produce a distinct Hercules. The slightly faster long bar model of \cite{HB18} produces Hercules, but not the low velocity moving groups, and the hat is too strong. It is worth noting that both of the long slow bar models are consistent within error with the recent measurements of the bar pattern speed \citep[e.g.][]{SSE19,Clarke+19}, depending on the choice of local circular frequency.

However, we do not expect the bar only models to produce a perfect representation of Solar neighbourhood kinematics because spiral structure will have some effect regardless of its nature, and so will the phase wrapping known to be occurring in the disc \citep[e.g.][]{Antoja+18} regardless of whether it originates from transient spiral arms \citep{HHBKG18}, or an external perturbation \citep[e.g.][]{LJGG-CB18}. 

\subsection{Density Wave spirals}
\label{wave}
We do not perform a detailed comparison of the different theories of spiral structure alone \citep[e.g. see][for a review of spiral structure]{DB14}, which was the focus of the recent work of \cite{Sw+18}. Instead, we show the effect of the combination of both a bar and spiral structure, for both the classical quasi-stationary density wave fixed pattern speed spiral of \cite{LS64} in this section, and the corotating winding arm as seen e.g. in \cite{GKC12} in Section \ref{corot} for bars of varying lengths and pattern speeds. We do not focus on making the best possible reproduction of the Solar neighbourhood kinematics, but merely show that the combination of bar and spiral leads to significant deviations from the symmetric resonances. 

A density wave like spiral, with a fixed pattern speed will have resonances at fixed radii in the same way as the bar described above. However, the spiral wave is expected to be slower than the bar, and thus the 4:1 ILR (blue) and the CR (red) of the spiral are the most likely to affect the velocity distribution in the Solar neighbourhood, whereas the OLRs will affect stars further from the Galactic centre.

We performed the action-angle modelling for an $N=2$ and an $N=4$ density wave like spiral arm with pitch angle 12 deg \citep[e.g. following the average measured angle of the Perseus arm;][]{V15}, and with a range of pattern speeds from $\Omega_{\mathrm{sp}}=0.2\times\Omega_0$ to $1.2\times\Omega_0$ in combination with each of the three bar models described above. There are too many plots to display within the paper but we would be happy to provide any model upon request. Figure \ref{1p8DW4} shows an example from the $N=4$ density wave spiral combined with a short fast bar for spiral pattern speeds $\Omega_{\mathrm{sp}}=0.56\times\Omega_0$ (upper), and $\Omega_{\mathrm{sp}}=0.91\times\Omega_0$ (lower).

The upper panel of Figure \ref{1p8DW4} shows Model D, the model with $\Omega_{\mathrm{sp}}=0.56\times\Omega_0$, for which the 4:1 ILR (blue) lies around $L_z=1.1$, corresponding to the ridge at the top of the main velocity distribution. The orange, yellow and magenta lines are the bar 4:1, 2:1 and 1:1 OLR respectively. The addition of the spiral density wave to the short bar model causes arches in the $L_z-\theta_{\phi}$ plane, which are qualitatively similar to the split between the moving groups in the $Gaia$ data, and the $\theta_{\mathrm{R}}-J_{\mathrm{R}}$ plane shows similar features around $\theta_{\mathrm{R}}=0$ and 2$\pi$. More interestingly, the second component of the Hercules like feature is stronger than in the bar only case, becoming more consistent with the data. The distribution as a whole does not have the correct vertex deviation, but otherwise it is a reasonably good match to the $Gaia$ data considering the simplicity of the model. While there is a strong edge to the velocity distribution around $L_z=1.1$ in the $L_z-\theta_{\phi}$ plane, the ridge does not extend to high $J_{\mathrm{R}}$ as is seen in the data, and the hat is not as distinct. We selected this combination of pattern speed for the bar and spiral arms because it was qualitatively the best fitting combination of a bar and density wave spiral from the sampled parameters when comparing the action-angle model to the data.

The lower panel of Figure \ref{1p8DW4} shows Model E, a model chosen to mimic the model from \cite{M-MPPV19}, where the ridge structure in the $R_{\mathrm{G}}-v_{\phi}$ plane was proposed to arise solely from a combination of bar and spiral resonances. The model is similar to the above panel except in this model the bar length is $R_{\mathrm{b}}=3.2$ kpc, and has pattern speed $\Omega_{\mathrm{b}}=1.82\times \Omega_0$. The spiral has a pattern speed of $\Omega_{\mathrm{sp}}=0.91\times\Omega_0$ and pitch angle 15.5 deg. Note that we do not have the same potential or the full parameters of their model. Hence, this will be only a qualitative reproduction based on the relevant spiral pattern speed, but the resonances should fall in the same area of the disc. Unlike the upper panel there is very little structure outside the bar OLR region. The spiral CR (red) creates a small amount of arching around $L_z=1$, but it is very weak compared to the data. 

\subsection{Corotating recurrent arms}
\label{corot}
The origin and processes behind the transient winding arms commonly seen in $N$-body simulations are still under debate, i.e. whether they arise from a series of overlapping transient modes \citep[e.g.][]{QDBMC11,CQ12,SC14}, or whether they are fully corotating recurrent arms which arise through a non-linear process similar to swing amplification \citep[e.g.][]{WBS11,GKC12,GKC12-2} which is difficult to explain via the superposition of modes \citep[e.g.][]{GKC12,Kumamoto+Noguchi16}. However, for this example we choose to reproduce the corotating recurrent arms \citep[e.g. as shown in][]{GKC12}, where the arm corotates with the stars at all radii, and does not have a fixed pattern speed. Note that this is a similar but not identical process to the classic swing amplification model proposed by \cite{T81}, and our model is not designed to reproduce a swing amplified arm as suggested in \cite{Sw+18}.

\begin{figure}
\centering
\includegraphics[width=\hsize]{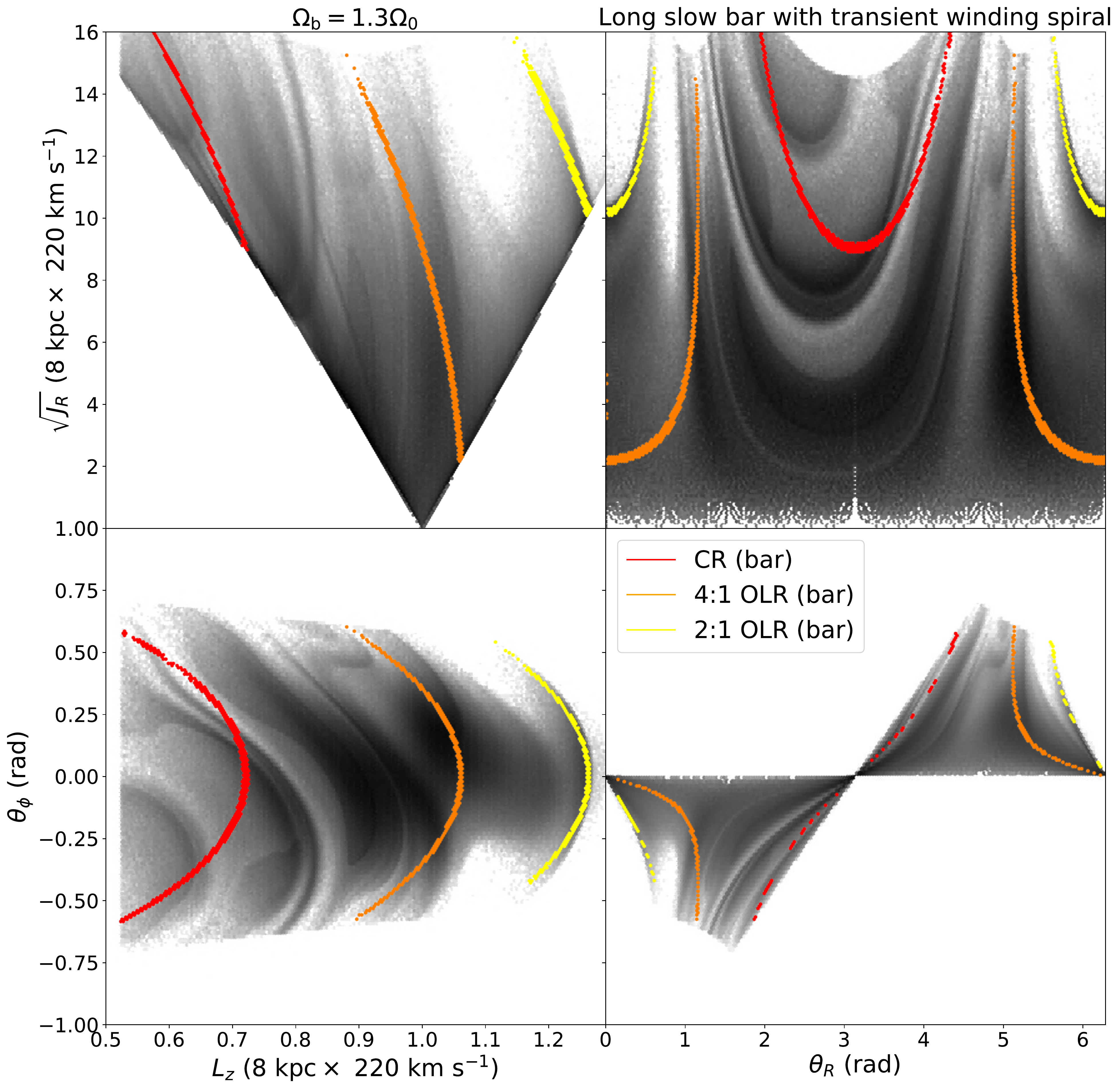}
\includegraphics[width=\hsize]{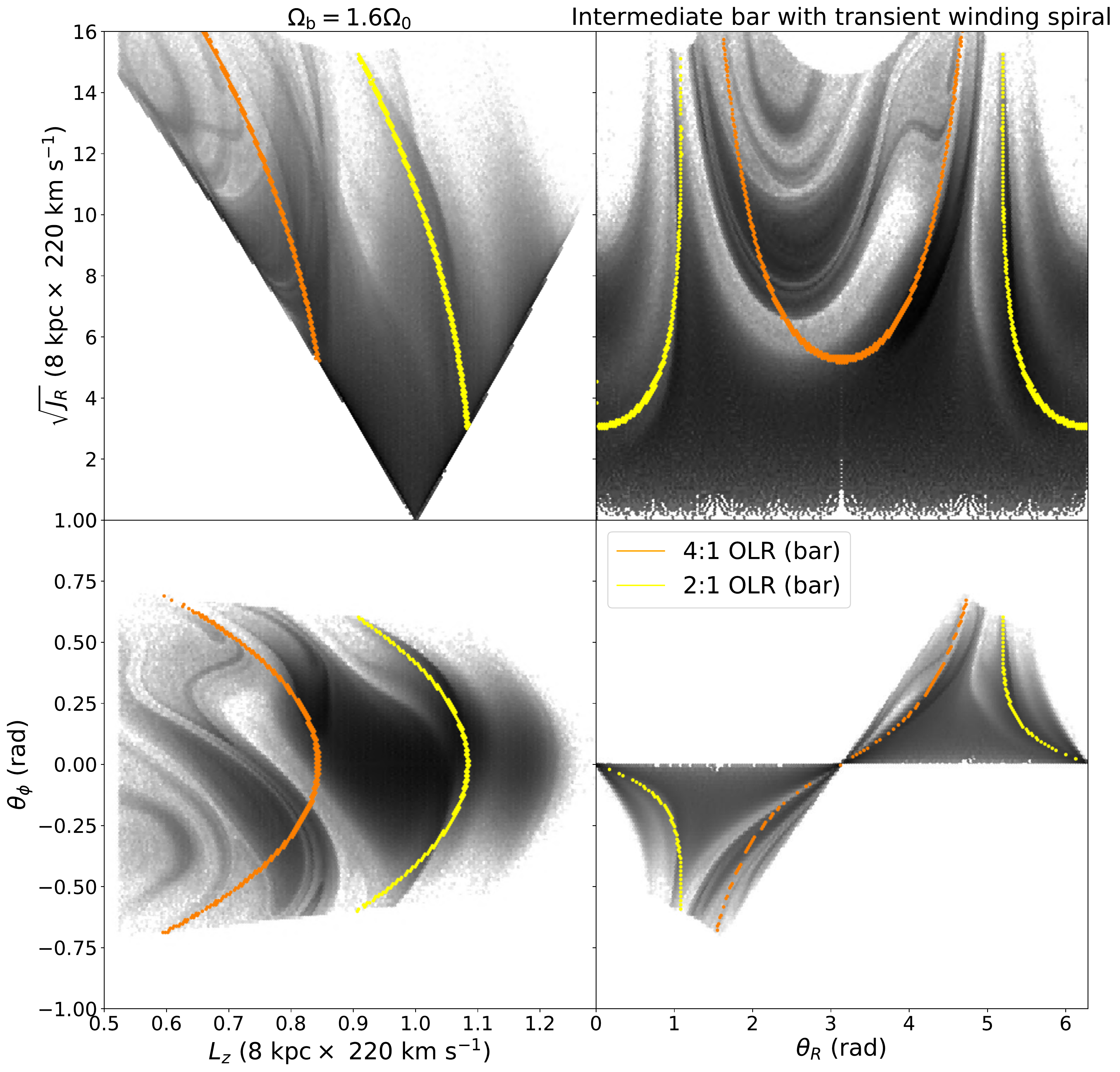}
\caption{Action-angle distribution for Model F, a long slow bar with $\Omega_{\mathrm{b}}=1.3\times\Omega_0$ (upper) and Model G, a bar with $\Omega_{\mathrm{b}}=1.6\times\Omega_0$ (lower), both combined with transient winding spirals. The spiral arms in these models corotate at all radii, and thus do not have resonance criteria to plot.}
\label{TW1}
\end{figure}

\begin{figure}
\centering
\includegraphics[width=\hsize]{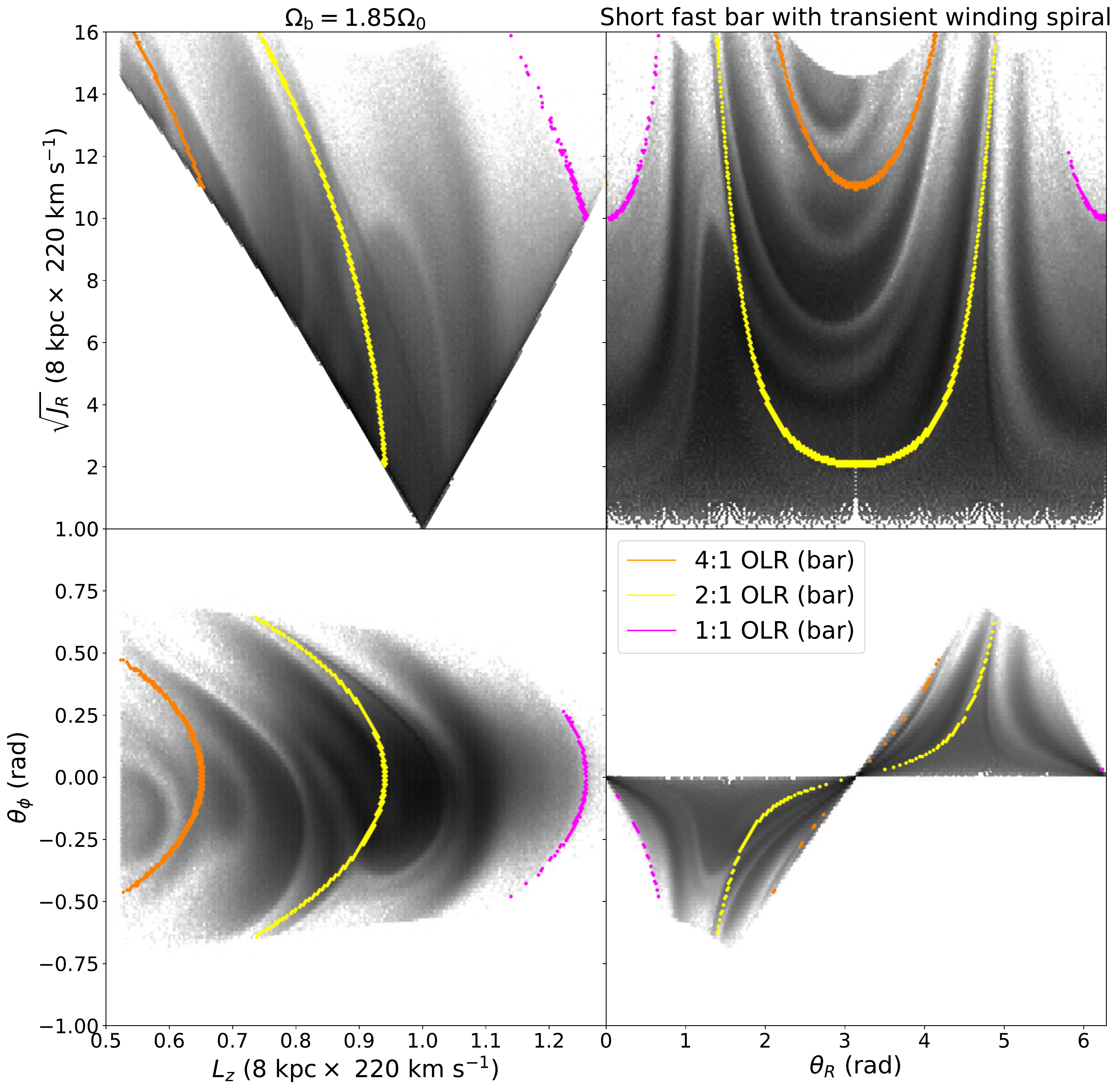}
\caption{Action-angle distribution for Model H, the short fast bar combined with transient winding spirals.}
\label{fastwinding}
\end{figure}

We again perform a series of action-angle models for a transient winding arm using the method described in Section \ref{models}, combined with a bar with varying length and pattern speed from $3<R_{\mathrm{b}}<5$ kpc, and $1.0\leq\Omega_{\mathrm{b}}\leq2.0\times\Omega_0$. The parameters of the spiral remain the same across models, with $N=2$, $R_{s}=0.3$ and $C_n=1$. We increase the pitch angle at the peak of the density enhancement from 12 degrees as assumed in \cite{HHBKG18} to 25 degrees which better matches the value found in \cite{GKC13}. This increases the strength of the perturbation on the Solar neighbourhood kinematics for an equivalent peak amplitude. The shape of the features are qualitatively the same, but more pronounced for the higher pitch angle at peak density.

The top panel of Figure \ref{TW1} shows Model F, the combination of a long slow bar with pattern speed $\Omega_{\mathrm{b}}=1.3\times\Omega_0$, and a series of three winding transient arms described above which peak at $t=-460$, $t=-230$ and $t=0$ Myr in the past \citep[e.g. the model from][but with a 25 deg pitch angle at the peak]{HHBKG18}. This places the Sun in the interarm region similar to the Milky Way. The model clearly shows the a hat like feature at high $L_z$ caused by the OLR which is similar to the observations from $Gaia$, and the central area of the velocity distribution contains three separate regions qualitatively similar to the Hyades+Pleiades, Coma Berenices and Sirius moving groups. There is also a distinct Hercules like feature, which is divided into separate substructures as seen in the data. The combination of the long slow bar and winding transient spiral structure naturally leads to the vertex deviation observed in the $L_z$ - $\theta_{\phi}$ plane, and the asymmetry around $\theta_{\mathrm{R}}=\pi$ in the $\theta_{\mathrm{R}}$ - $\sqrt{J_{\mathrm{R}}}$ plane. It is worth noting that while we selected $\Omega_{\mathrm{b}}=1.3\times\Omega_0$ as the `best' qualitative reproduction of the Solar neighbourhood kinematics for this model, the features are similar over an approximate range of $\Omega_{\mathrm{b}}\sim1.26-1.36\times\Omega_0$. The bar pattern speed found by \cite{MFSWG18} falls within this range, and the features from the bar are consistent across both models. However, the addition of spiral structure adds additional features and strengthens the Hercules like feature.

The lower panel of Figure \ref{TW1} shows Model G, the same combination of potentials, but with a bar pattern speed of $\Omega_{\mathrm{b}}=1.6\times\Omega_0$, which shows a surprisingly good reproduction of the Solar neighborhood kinematics. The Hercules like division is both the cleanest of any combination of parameters which were explored, and both matches the observed angle and location in $L_z$. The Hercules like feature is comprised of multiple components, and is asymmetric in $\theta_{\mathrm{R}}$. It results from the 4:1 OLR of the bar, and the transient winding spiral arms. The 2:1 OLR causes the sharp edge of the high velocity area of the distribution. The OLR occurs at lower $L_z$ and makes a qualitatively similar feature to Sirius. There is then a small arch above this, which matches the hat in the data well. The vertex deviation of the distribution matches the observations, but there is no clear separation into low velocity moving groups. 

Figure \ref{fastwinding} shows Model H, the combination of a short fast $m=2$ bar potential with $\Omega_{\mathrm{b}}=1.85\times\Omega_0$ with a series of transient winding spiral arms. These transient arms have the same parameters, but their spacing has been altered such that the peaks occur slightly more frequently at 210 Myr years apart instead of 230 Myr apart. Figure \ref{fastwinding} shows a multiple component Hercules like feature, with a prominent main ridge, a small ridge between it and the Hyades/Pleiades like feature and smaller ridges at lower $L_z$. The model contains a small hat feature resulting from the 1:1 OLR of the bar, the correct vertex deviation, along with the strong feature resembling Sirius, although this extends further from $\theta_{\phi}=0$ than is seen in the data. 

\begin{figure}
\centering
\includegraphics[width=\hsize]{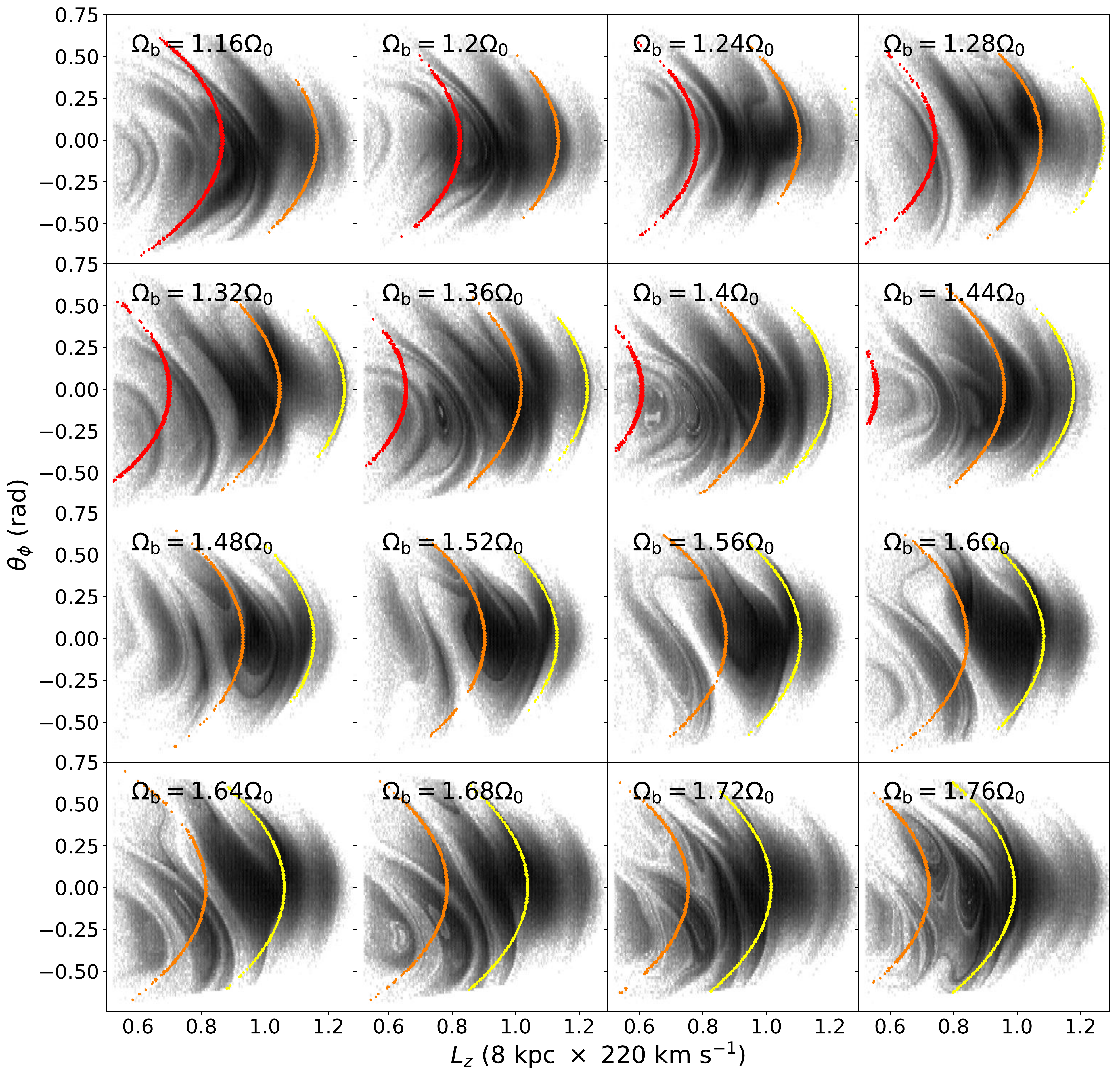}
\caption{$L_z-\theta_{\phi}$ distribution for a transient winding spiral model combined with a bar model with a range of pattern speeds from $\Omega_{\mathrm{b}}=1.16\times\Omega_0$ to $\Omega_{\mathrm{b}}=1.76\times\Omega_0$, and bar radius which decreases with increasing pattern speed. As with earlier figures, the colored lines mark the location of the CR (red), 4:1 OLR (orange) and the 2:1 OLR (yellow).}
\label{the16}
\end{figure}

Again, the full range of pattern speeds are available upon request. We do not claim that any of the above solutions are the correct representation of the Milky Way's non-axisymetric structure, but merely that there are a range of solutions that are consistent with the kinematics of the Solar neighbourhood. When exploring a bar only model \citep[e.g.][]{D00,HB18,MFSWG18} the features are distinct, both in action-angle space and the $v_{\mathrm{R}}-v_{\phi}$ plane, and there are a very limited number of pattern speeds which are able to account for Hercules, or other features such as the hat. Similarly, when exploring spiral only models such as was done in \cite{Sw+18}, the features in $v_{\mathrm{R}}-v_{\phi}$ and action-angle space are clean and distinct from one another. However, when analysing a combination of bar and spiral structure, the overlapping or coupling of multiple resonances and phase mixing make the picture a lot less clear.

To illustrate this, Figure \ref{the16} shows the $L_z-\theta_{\phi}$ plane for the same model as Figure \ref{TW1}, but with a range of pattern speeds from $\Omega_{\mathrm{b}}=1.16\times\Omega_0$ to $\Omega_{\mathrm{b}}=1.76\times\Omega_0$, and a bar length which decreases with pattern speed. As with earlier Figures, the colored lines show the CR (red), 4:1 OLR (orange) and the 2:1 OLR (yellow) for each bar pattern speed. All other model parameters (including the spiral model) remain the same between panels. The panels with $\Omega_{\mathrm{b}}=1.28\times\Omega_0$ and $\Omega_{\mathrm{b}}=1.6\times\Omega_0$ are the closest to the data, but a large range of pattern speeds show divisions around the location of Hercules, and arch features at high $L_z$. Also note that while some features are broad, some features are very fine, in agreement with the data. 

We also note that the inclusion of transient spiral arms to either the short fast bar or the long slow bar can be tailored to reproduce the structure which is otherwise missing. E.g. adding spirals to a long slow bar can reproduce a distinct Hercules, which it does not do alone, and adding spirals to a short fast bar can reproduce the structure at higher $L_z$, which it does not do alone. We do not expect any of these models to be a perfect reproduction of the Solar neighbourhood because this is a drastic simplification of the complex dynamical behaviour ongoing in the Milky Way disc, and we have not attempted to fit the timescale of the phase mixing.

\subsection{The $R_{\mathrm{G}}-v_{\phi}$ plane}
\label{rvp}
In addition to the kinematics of the Solar neighbourhood, we also compare the models further from the Solar circle. Thus, in this section, we show the $R_{\mathrm{G}}-v_{\phi}$ plane for the models described above, and compare them with the $Gaia$ data, both in terms of stellar number density as we did previously in \cite{HHBKG18}, and also coloured by $v_{\mathrm{R}}$ as shown in \cite{Fragkoudi+19}.

\begin{figure*}
    \centering
    \includegraphics[width=\hsize]{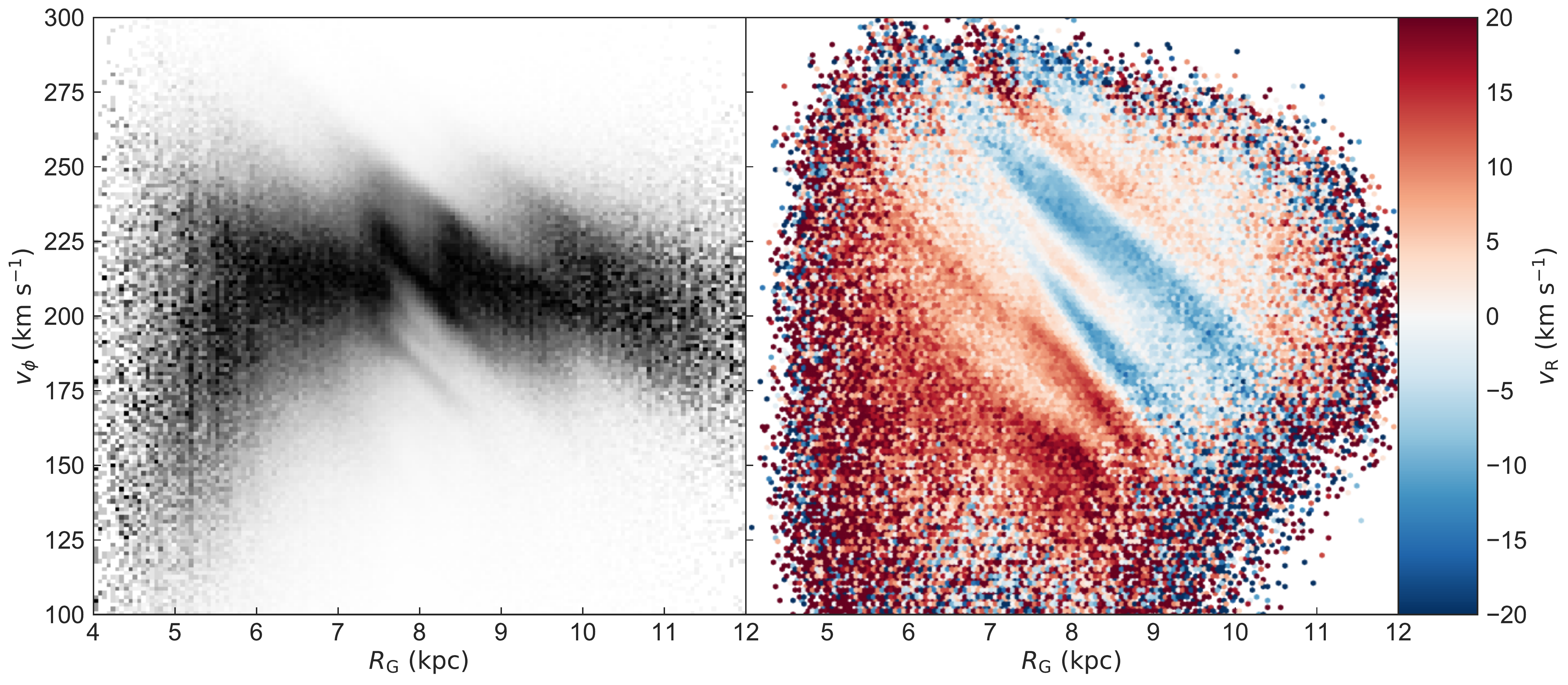}
    \caption{$v_{\phi}$ as a function of Galactocentric radius $R_{\mathrm{G}}$, colored by normalised number density (left), and $v_{\mathrm{R}}$ (km s$^{-1}$ (right), for the $Gaia$ DR2 data.}
    \label{RRGaia}
\end{figure*}

Figure \ref{RRGaia} shows $v_{\phi}$ as a function of Galactocentric radius $R_{\mathrm{G}}$, colored by normalised number density (left column), and $v_{\mathrm{R}}$ (right column), for the $Gaia$ DR2 data. Unfortunately, even with the unparalleled wealth of data from $Gaia$ DR2, we are still only able to trace the ridges a few kpc from the Solar circle. The vertex deviation of the velocity distribution means the moving groups are not fully distinct in $v_{\phi}$. However, some moving groups do dominate some band of the $v_{\phi}-R$ plane. For example, the multiple Hercules streams are moving outwards (red) as expected with the uppermost of the two dark red feature being the main peak of Hercules, and the paler feature just above being the smaller peak at slightly higher $v_{\phi}$. The horn feature moving inwards (blue) is just `above' Hercules \citep[as explored in more detail in][]{Fragkoudi+19}. The wide inwards moving (blue) band of stars is mostly Sirius. The outwards moving band just `below' Sirius is mostly the Hyades, but Coma Berenices lies on the transition from the Sirius band to the Hyades band, and the Pleiades also lies along the transition from the Hyades band to the horn. It is not quite so simple to trace the Coma Berenices or Pleiades moving groups in this projection because they have approximately $\bar{v}_{\mathrm{R}}=0$ and merely decrease the amplitude of the $v_{\mathrm{R}}$ signal of the adjacent bands. Thus we do not add explicit labels to this projection.

\begin{figure}
    \centering
    \includegraphics[width=\hsize]{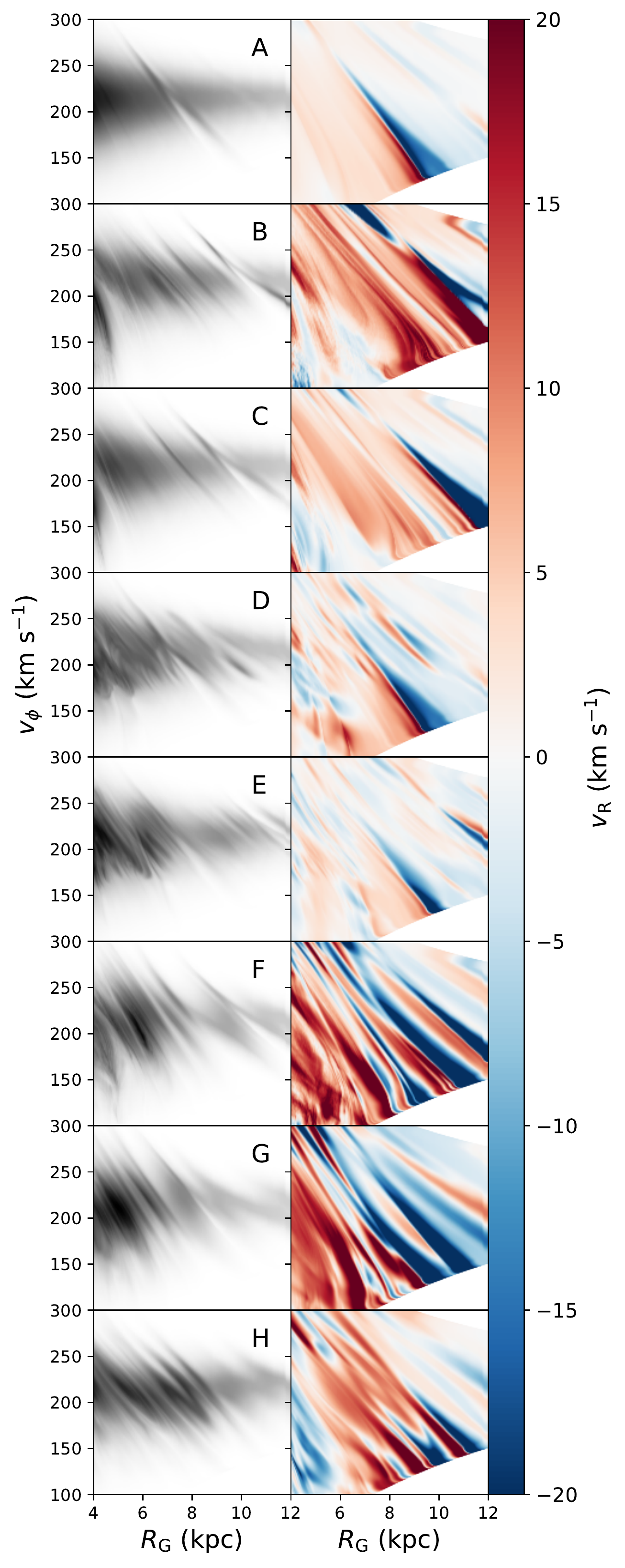}
    \caption{$v_{\phi}$ as a function of Galactocentric radius $R_{\mathrm{G}}$, colored by number density (left column), and $v_{\mathrm{R}}$ (km s$^{-1}$ (right column), for models A, B, C, D, E, F, G \& H. (top to bottom). Their parameters are given in Table \ref{params}.}
    \label{RadialRidges}
\end{figure}

Figure \ref{RadialRidges} shows the logarithmic number density (left column), and the mean radial velocity $v_{\mathrm{R}}$ (right column) for the previously shown models. Note that the mean $v_{\mathrm{R}}$ shown in the right hand panels is not weighted by number density, so features may appear strong even when there are very few stars. Thus, this should not be taken to indicate that the models have an abnormally large velocity dispersion.

The short fast bar in model A (top row) shows the single OLR feature around the Solar neighbourhood as expected, with the red outward moving stars corresponding to the Hercules stream, and the blue inwards moving stars corresponding to the horn \citep[the orbital behaviour is explored in detail in][]{Fragkoudi+19}. There are minor effects present at other radii including a weak red-blue feature around the 1:1 OLR, and a small secondary Hercules-like ridge, but the distribution is mostly smooth away from the 2:1 OLR.

The long slow bar in model B (second row) shows a lot more structure away from the 2:1 OLR which occurs further out in the Galaxy and is identifiable from the sharp diagonal transition from red to blue. The 1:1 OLR makes another weaker transition in the top right corner of the panel. There are multiple red `stripes' resembling Hercules, but the model lacks the alternating inwards and outwards moving groups of stars which are present in the data, with very little blue. In particular, there is no blue horn-like feature.

The slightly faster long bar model C (3rd row) shows very similar behaviour to model B, which is unsurprising considering they both contain an $m=4$ component. However, the slightly faster pattern speed puts the OLR nearer the Solar circle, and the lower bar strength reduces the relative strength of the inwards and outwards mean velocities. This lower strength allows a small blue horn-like feature to become apparent above the Hercules-like feature \citep[for a comparison of the effect of the strength of the $m=4$ component see][]{HB18}.

The short fast bar + four armed density wave model D (4th row) more closely resembles model A. E.g. the 2:1 OLR is the only strong feature in the Solar neighbourhood. The combination of the bar and spiral resonances create multiple ridges in the number density $R_{\mathrm{G}}-v_{\phi}$ plane, which is qualitatively consistent with the data. However, when colored by $v_{\mathrm{R}}$ there is significantly less variation than is seen in the data. There is a red blue feature around the 4:1 ILR of the density wave, but weaker than seen in the data.

The short fast bar + four armed density wave spiral with pattern speeds following \cite{M-MPPV19}, model E (fifth row) shows very similar behaviour to model D. There is ridge structure in the density panel, essentially following the same distribution as in \cite{M-MPPV19}, but the $v_{\mathrm{R}}$ panel shows only weak structure outside the OLR region, including at the spiral resonances.

The long slow bar + transient winding spirals model F (6th row) shows more strongly alternating inwards and outwards moving stars, which is to be expected from a phase mixing model \citep[e.g. see][]{Khanna+19}. That these overdensity
ridges at approximately constant $L_z = R \times v_{\phi}$ are related to regions of predominantly outward or inward motion was first shown by Trick et al. (2019). The OLR again occurs further out in the disc, but the Solar neighbourhood contains multiple Hercules like features, a blue Sirius like feature, and then the alternating red and blue in the hat like feature caused by the 2:1 OLR, which is a good match to the data. There is also a very thin blue horn like feature, but the Hyades like feature is too strong compared to the data.

The bar with $\Omega_{\mathrm{b}}=1.6\times\Omega_0$ + transient winding spirals model G (7th row) shows a similar mix of inwards and outwards moving stars consistent with phase mixing, but the amplitude of the variation is much higher. Again, there are multiple Hercules like features, and multiple blue horn like features are present, as are the red Hyades and the blue Sirius like features. However, the inwards and outwards moving components of the hat like feature are opposite to what is observed, with the inwards moving stars occurring at lower $v_{\phi}$ than the outwards moving stars, contrary to the data.

The short fast bar + transient spirals model H (bottom row) again shows some alternating ridges. However, interestingly the amplitude of the mean $v_{\mathrm{R}}$ is significantly lower outside the area affected by the bar OLR. This appears to be consistent with models F \& G, but the effect is stronger here. While there are distinct multiple Hercules like features, and a small horn like feature, there are considerably too many red outwards moving features which are not consistent with the data. 

In general, the models combining the bar with transient spiral structure are qualitatively a better fit to the $Gaia$ data, with strong alternating inwards and outwards moving groups of stars. The bar only models affect too narrow a region of the phase space with various resonances, and the density wave spirals models do not leave a strong signal in $v_{\mathrm{R}}$ with the spiral induced features being approximately symmetrical around $v_{\mathrm{R}}=0$ (e.g. see Figure \ref{1p8DW4}). The three transient spiral phase mixing models are qualitatively consistent with the $Gaia$ data, but none are an excellent match. This is unsurprising because no attempt has been made to match the (currently unknown) timescale of the phase mixing event, and the backwards integration models described above in Section \ref{models} are evaluated only once at a single timestep.

\subsection{Test particle simulations}\label{TPM}
We also run high resolution test particle models for models F \& H (selected as the two closest to the data) in order to explore how the phase mixing features change with time. This also serves as a useful check that the results from the simple models constructed with the backwards integration method are consistent with the test particle simulations.

The initial conditions for the test particle models are sampled using \texttt{galpy}'s \texttt{quasiisothermaldf}, a distribution function adapted from \cite{B10} which is expressed in terms of action-angle variables. The distribution function is initialized with a radial scale length of $R_0/3$, a local radial and vertical velocity dispersion of $0.15 \times V_{\mathrm{c}}(R_0)$ and $0.075\times V_{\mathrm{c}}(R_0)$, respectively, and a radial scale length of $R_0$ for the velocity dispersions. Action-angles are calculated using the Staeckel approximation via \texttt{galpy}'s \texttt{actionAngleStaeckel}. The potential is evaluated with \texttt{MWPotential2014}. $2 \times 10^8$ position and velocity samples are generated between radii of 1 and 13 kpc, and are integrated forward in time for 7 Gyr using \texttt{MWPotential2014} to allow the disc to reach equilibrium prior to the model integration. The bar and spiral potentials used for models F and H are the 3D versions of those given above, and both models are then integrated forwards for 10 bar periods using the \texttt{MWPotential2014} potential.

\begin{figure*}
    \centering
    \includegraphics[clip, trim=2cm 0cm 2cm 0cm,width=\hsize]{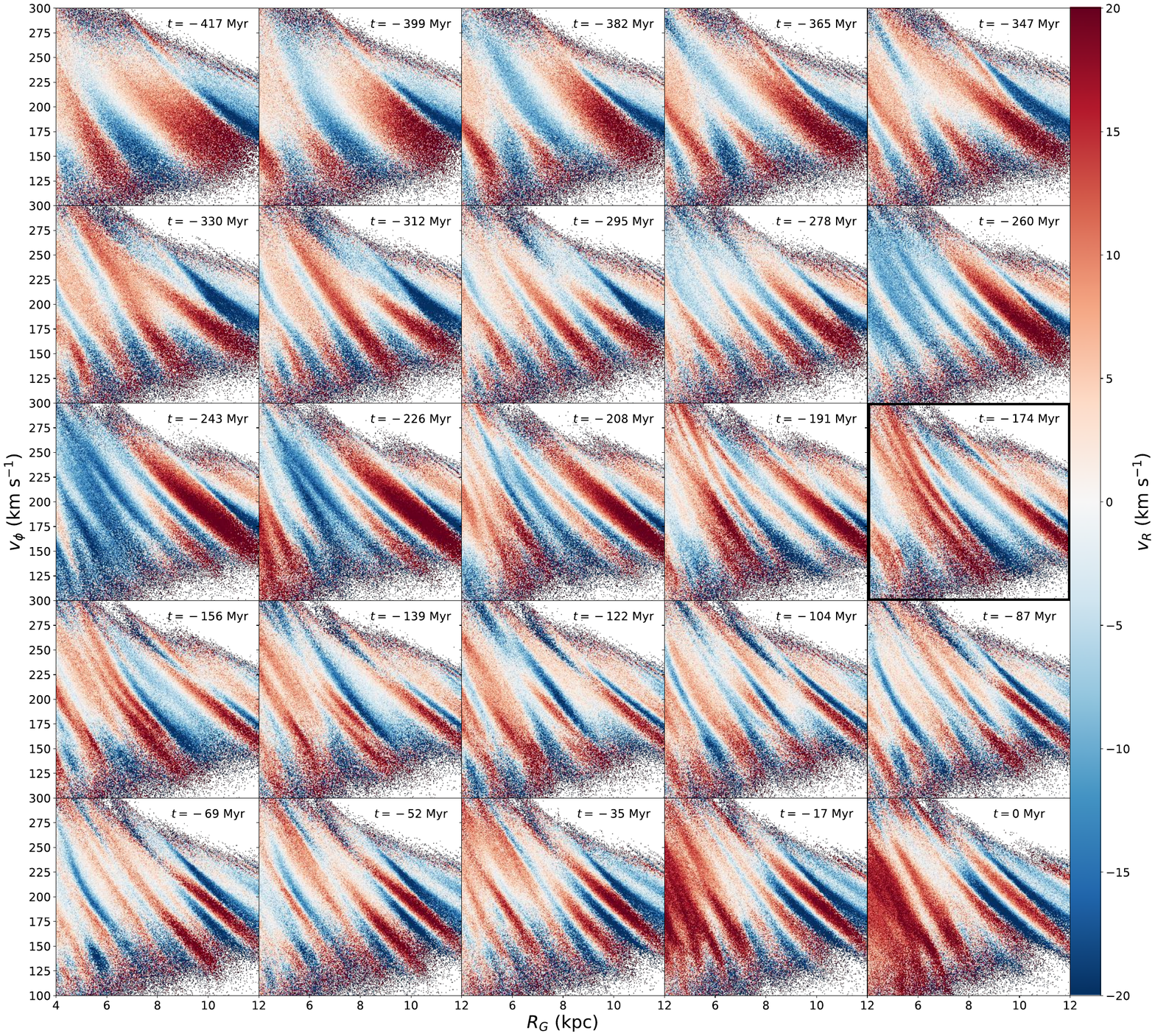}
    \caption{$v_{\phi}$ as a function of Galactocentric radius $R_{\mathrm{G}}$, colored by mean $v_{\mathrm{R}}$ for a series of timesteps from $t=-417$ Myr to $t=0$ (present day) for the test particle realisation of Model F. The best fitting panel at $t=-174$ Myr is outlined.}
    \label{MB1}
\end{figure*}

Figure \ref{MB1} shows the time evolution of the $R_{\mathrm{G}}-v_{\phi}$ plane for the test particle realisation of model F, i.e. a long slow bar combined with transient winding spirals colored by mean $v_{\mathrm{R}}$. The time is given in the top corner of each panel from $t=-417$ to $t=0$ Myr, with $t=0$ being the present day. The spirals peak at $t=-460$, $t=-230$ and $t=0$ Myr. The simulation has the same overall timescale as the backwards integration models, as we integrate forwards from $t=-1.72$ Gyr to 0 (after the relaxation phase). All panels are taken at a Galactic azimuth such that the bar is at 25 deg from the line of sight from the Galactic centre to the observer, and the particles are selected in a wedge of $\pm15$ deg around the Galactic centre-Sun line. The panel at $t=0$ is equivalent to the right hand panel in the 6th row (F) of Figure \ref{RadialRidges}, and they show excellent agreement, validating the backwards integration $R_{\mathrm{G}}-v_{\phi}$ planes. The test particle realisation shows more blurring between features owing both to the finite resolution and the sampling of a wedge of particles, whereas the backward integration model is evaluated along an exact line from the Galactic centre to anti-centre. We remind the reader that as with Figure \ref{RadialRidges}, the mean $v_{\mathrm{R}}$ shown in the right hand panels is not weighted by number density, so features may appear strong even when there are very few stars. Thus, this should not be taken to indicate that the models have an abnormally large velocity dispersion.

The bar OLR is visible in every panel, with the transition from outwards to inwards moving features occuring around 10-12 kpc. The strength of the feature varies between timesteps, but never gets fully obscured by the effects of the phase mixing from the transient spiral arms. The data in Figure \ref{RRGaia} is very noisy in the hat region, but the trend of a slight inwards moving component at high $v_{\phi}$, with the outwards moving component between it and Sirius is consistent with the tip of the OLR feature. While the $t=0$ panel is a reasonable match to the data, with a strong Hercules stream, and also Sirius and Hyades like features, the panels for $t=-174$ (outlined) and $t=-156$ show a significantly better reproduction of the Solar neighbourhood kinematics. For example, around $R_{\mathrm{G}}=8$ kpc there is a double Hercules like outward moving feature, a tapering inwards moving feature resembling the horn, a thin Hyades like feature and a broad Sirius like feature. However, the red/blue transition in the hat is stronger than in the data.

In general, the inwards moving part (blue) of the OLR feature maintains a similar shape accross most panels, where it is broad at large $R_{\mathrm{G}}$ and low $v_{\phi}$ and becomes narrower towards lower $R_{\mathrm{G}}$ and high $v_{\phi}$. The strength and extent of the blue OLR feature does fluctuate between panels, but overall the feature remains qualitatively the same, even when it becomes very thin. However, the outwards moving (red) part of the OLR feature shows significant variation over the timescale displayed in Figure \ref{MB1}. In some panels, e.g. $t=0$, the outwards moving part of the OLR feature closely follows the inwards moving feature, while maintaining an approximately constant width across the covered $R_{\mathrm{G}}$ and $v_{\phi}$ range. However, in other panels, the outwards moving part of the OLR feature splits significantly from the inwards moving part, either leaving a gap between them at higher $v_{\phi}$  (e.g. $t=-295$ Myr), or leaving a thin strip of outwards moving stars next to the inwards moving stars, while the stronger feature splits (e.g. $t=-347$ Myr). This behaviour is qualitatively similar to the Hercules stream observed in the Solar neighbourhood albeit at a different $R_{\mathrm{G}}$ and $v_{\phi}$.

For example, Figure \ref{RRGaia} shows that the horn becomes broader at higher $R_{\mathrm{G}}$, lower $v_{\phi}$ and while the strongest part of the Hercules stream itself splits from the horn, there is a smaller strip of stars which remain next to the horn. These separate groups are also visible as two of the multiple components of Hercules in the $v_{\mathrm{R}}-v_{\phi}$ or $L_z-\theta_{\phi}$ plane, e.g. the strong main peak of Hercules splits from the Horn, and the smaller peak at slightly higher $v_{\phi}$ remains next to the horn. And while the long slow bar OLR cannot be responsible for this behaviour in the Solar neighbourhood, it is conceivable that the same effect can happen around a difference resonance, e.g. the CR or the 4:1 OLR. This interpretation, if correct, would mean that the smaller peak is the true resonance feature, whereas the larger peak of Hercules originates from the coupling of a resonance and the phase mixing.

\begin{figure*}
    \centering
    \includegraphics[width=\hsize]{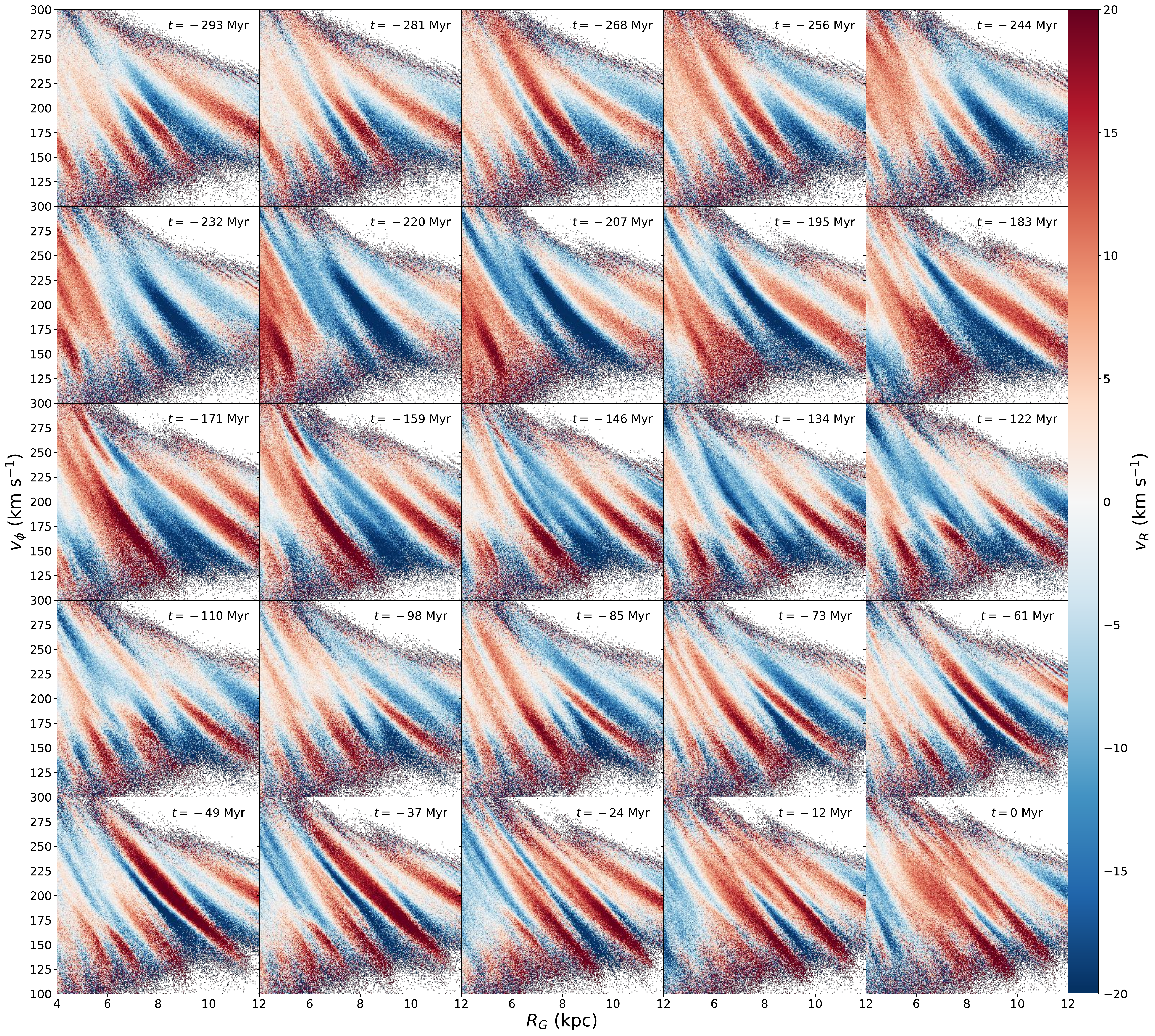}
    \caption{$v_{\phi}$ as a function of Galactocentric radius $R_{\mathrm{G}}$, colored by mean $v_{\mathrm{R}}$ for a series of timesteps from $t=-293$ Gyr to $t=0$ Gyr (present day) for the test particle realisation of Model H.}
    \label{MB2}
\end{figure*}

Figure \ref{MB2} shows the time evolution of the $R_{\mathrm{G}}-v_{\phi}$ plane for the test particle realisation of model H, i.e. a short fast bar combined with transient winding spirals, where the arms peak at $t=-420$, $t=-210$ and $t=0$ Myr. In this model, the OLR feature which occurs around 7-9 kpc is no longer clear at every timestep, despite the constant bar angle. However, multiple panels show a clear horn like feature, and Hercules like feature, with a tapering inwards moving group of stars, and a strong outwards moving group at slightly lower $v_{\phi}$ resulting from the 2:1 OLR. Note that this is similar to the findings of \cite{FBBPZ18} who observe a Hercules like feature slightly outside the OLR in approximately 50\% of the sampled time steps in their high resolution $N$-body simulations, depending on the complex interaction between the bar and transient spiral structure.

The panel at $t=-73$ Myr contains a strong horn like feature and a double Hercules like feature, and also a broad Sirius like feature. There is a thin strip of outwards moving stars qualitatively consistent with Hyades, but the amplitude of the $v_{\mathrm{R}}$ is much stronger than seen in the data. The panels where $t=-159, -146$ Myr contain similar behaviour as discussed above, where a small strip of outward moving stars remain next to the inwards moving feature, and then a stronger feature splits away. These two panels also contain a thin slightly outwards moving feature just outside the horn like feature which is qualitatively consistent with Hyades, and then a broad inwards moving feature just outside that which is consistent with Sirius. The hat like feature then contains the split outward and inward moving features, although the outward moving group is stronger than seen in the data. The panel at $t=0$ is a significantly worse fit to the data, with the majority of the stars in the Solar neighbourhood moving outwards. There is a small horn like feature, and a strong Hercules like feature, but the rest of the distribution is a poor match to the data. Some of the phase mixing ridges survive across 100 Myr, whereas others only appear in one or two panels, i.e. only around 25 Myr, especially when interacting with the OLR. The trend observed in Figure \ref{RadialRidges} where the phase mixing ridge features were weaker outside the bar 2:1 OLR does not appear to be consistent across panels, and thus cannot be reliably used to infer the location of the OLR.

Further work is necessary to fully explore the timescale of the phase mixing, both regarding how long the features last from a single perturbation when combined with a barred potential, and also the effects of how frequently the perturbations occur, but we defer this to a future work.

\section{Summary}\label{summary}
In this work we have investigated the effects of the combination of bar and spiral potentials on the action-angle distribution of stars in the Solar neighbourhood. 

In general, in both the action-angle planes, and the $R_{\mathrm{G}}-v_{\phi}$ planes we find it is relatively easy to qualitatively reproduce observations via a combination of a bar and transient winding spiral arms. This holds true for a variety of bar lengths and pattern speeds when considering only the Solar neighbourhood kinematics. With respect to the spiral structure, quasi-stationary density wave spirals make significantly less impact on the explored kinematics, even when resonances fall in the right area of phase space, and bar only models simply cannot account for all the observed structure, which is not surprising.

Both the long slow bar and the short fast bar combined with transient spiral structure make a good fit to both the action-angle distribution and the $R_{\mathrm{G}}-v_{\phi}$ plane. The phase mixing effects from the transient spiral arms are highly time dependent, and when selecting a particular time step with respect to the phase mixing, we are able to reproduce the shape of the horn, and the multiple components of Hercules. However, we are not suggesting that this is the correct sequence of spiral structure, and it is likely that an even better fit to the data could be constructed with some other combination of transient spirals. Currently, the kinematic signatures in the Solar neighbourhood are more in favour of a shorter faster bar, whereas the more direct measurements of the bar length \citep[e.g.][]{WG13}, or pattern speed \citep[e.g.][]{SM15,Portail+17,SSE19,Clarke+19} are consistently in favour of a longer slower bar. In this work we have shown that the combination of a long slow bar and transient spiral structure can make a strong Hercules like feature, with a distinct gap between Hercules and the Hyades/Pleiades moving groups removing one of the key objections to a long slow bar.

Once we can trace the moving groups and ridges further across the disc we may be able to distinguish the origin of individual features. For example, the behaviour around the OLR is distinctive, and if the bar is long and slow we should be able to observe an OLR like feature further out in the disc, a radius at which we cannot currently sample reliably, but which will be accessible in the near future with $Gaia$ DR3 if not before. We may be seeing the edge of it in the hat, but the signal is weak. Currently there are too many models which can qualitatively reproduce Hercules and the other moving groups in the Solar neighbourhood to conclusively infer properties such as the bar pattern speed from Solar neighbourhood kinematics alone. However, the combination of a bar and transient winding spiral structure can reproduce the Solar neighbourhood moving groups, and the radially extended ridges seen in the $Gaia$ data.

\section*{Acknowledgements} J. Hunt is supported by a Dunlap Fellowship at the Dunlap Institute for Astronomy \& Astrophysics, funded through an endowment established by the Dunlap family and the University of Toronto. M. Bub and J. Bovy received support from the Natural Sciences and Engineering Research Council of Canada (NSERC; funding reference number RGPIN-2015-05235). J. T. Mackereth acknowledges support from the ERC Consolidator Grant funding scheme (project ASTEROCHRONOMETRY, G.A. n. 772293). D. Kawata acknowledges the support of the UK's Science \& Technology Facilities Council (STFC Grant ST/N000811/1 and ST/S000216/1). This project was developed in part at the 2018 NYC $Gaia$ Sprint, hosted by the Center for Computational Astrophysics of the Flatiron Institute in New York City. This project was developed in part at the 2019 Santa Barbara $Gaia$ Sprint, hosted by the Kavli Institute for Theoretical Physics at the University of California, Santa Barbara. This research was supported in part at KITP by the Heising-Simons Foundation and the National Science Foundation under Grant No. NSF PHY-1748958. Computations were performed on the Niagara supercomputer at the SciNet HPC Consortium \citep{Loken_2010}. SciNet is funded by: the Canada Foundation for Innovation; the Government of Ontario; Ontario Research Fund - Research Excellence; and the University of Toronto. This research made use of \texttt{astropy}, a community-developed core Python package for Astronomy \citep{Astropy}, \texttt{TOPCAT} \citep{Topcat}, and \texttt{galpy} \citep{B15}. This work has also made use of data from the European Space Agency (ESA) mission $Gaia$ (https://www.cosmos.esa.int/gaia), processed by the $Gaia$ Data Processing and Analysis Consortium (DPAC, https://www.cosmos.esa.int/web/gaia/dpac/consortium). Funding for the DPAC has been provided by national institutions, in particular the institutions participating in the $Gaia$ Multilateral Agreement.

\bibliographystyle{mn2e}
\bibliography{ref2}

\label{lastpage}
\end{document}